\documentclass[a4paper,11pt]{article}
\pdfoutput=1 
\usepackage{float}
\usepackage{tabulary}
\usepackage{verbatim}
\usepackage[dvipsnames]{xcolor}
\usepackage{textcomp}
\usepackage{gensymb}
\usepackage{url}		
\usepackage{subcaption}			
	
\usepackage{jcappub} 
\title{\boldmath Dark matter searches by the planned gamma-ray telescope GAMMA-400}

\author[a,1]{Andrey E. Egorov, \note{Corresponding author.}}
\author[a]{Nikolay P. Topchiev,}
\author[a,b]{Arkadiy M. Galper,}
\author[a]{Oleg D. Dalkarov,}
\author[a,b]{Alexey A. Leonov,}
\author[a]{Sergey I. Suchkov}
\author[b]{and Yuriy T. Yurkin}

\affiliation[a]{Nuclear Physics and Astrophysics Division, Lebedev Physical Institute, Leninskii prospect, 53, Moscow, Russia}
\affiliation[b]{National Research Nuclear University MEPhI, Kashirskoe highway, 31, Moscow, Russia}

\emailAdd{eae14@yandex.ru}

\abstract{Our paper reviews the planned space-based gamma-ray telescope GAMMA-400 and evaluates in details its opportunities in the field of dark matter (DM) indirect searches. We estimated the GAMMA-400 mean sensitivity to the diphoton DM annihilation cross section in the Galactic center for DM particle masses in the range of 1--500 GeV. We obtained the sensitivity gain at least by 1.2--1.5 times (depending on DM particle mass) with respect to the expected constraints from 12 years of observations by Fermi-LAT for the case of Einasto DM density profile. The joint analysis of the data from both telescopes may yield the gain up to 1.8--2.3 times. Thus the sensitivity reaches the level of annihilation cross section $\langle \sigma v \rangle_{\gamma\gamma}(m_\chi = 100~\mbox{GeV}) \approx 10^{-28}$ cm$^3$/s. This will allow us to test the hypothesized narrow lines predicted by specific DM models, particularly the recently proposed pseudo-Goldstone boson DM model. We also considered the decaying DM - in this case the joint analysis may yield the sensitivity gain up to 1.1--2.0 times reaching the level of DM lifetime $\tau_{\gamma\nu}(m_\chi = 100~\mbox{GeV}) \approx 2 \cdot 10^{29}$ s. We estimated the GAMMA-400 sensitivity to axion-like particle (ALP) parameters by a potential observation of the supernova explosion in the Local Group. This is very sensitive probe of ALPs reaching the level of ALP-photon coupling constant $g_{a\gamma} \sim 10^{-13}~\mbox{GeV}^{-1}$ for ALP masses $m_a \lesssim 1$ neV. We also calculated the sensitivity to ALPs by constraining the modulations in the spectra of the Galactic gamma-ray pulsars due to possible ALP-photon conversion. GAMMA-400 is expected to be more sensitive than the CAST helioscope for ALP masses $m_a \approx (1-10)$ neV reaching $g_{a\gamma}^{min} \approx 2 \cdot 10^{-11}~\mbox{GeV}^{-1}$. Other potentially interesting targets and candidates are briefly considered too.}

\keywords{dark matter experiments, dark matter theory, gamma ray experiments, axions}

\begin{document}
\maketitle
\flushbottom

\section{A brief overview of the GAMMA-400 project}
\label{sec:i}
GAMMA-400 (Gamma-ray Astronomical Multifunctional Modular Apparatus up to (initially) $\approx$ 400 GeV) is the Russian space-based pair-converting calorimetric gamma-ray telescope planned for a launch after 2026 (e.g. \cite{2017PAN....80.1141G}). It is expected to continue gamma-ray astronomy development after Fermi-LAT and other instruments using better characteristics. One of the key mission objectives is indeed DM indirect searches, where the gamma-ray range is historically the most promising channel (e.g. \cite{2018RPPh...81f6201R}). Our paper describes quantitatively the main directions in the field for GAMMA-400, where it can bring a significant progress. Mainly this refers to the two most popular DM candidates - WIMPs and ALPs. But other candidates are being considered too, especially in section \ref{sec:l}. Thus section \ref{sec:l} is dedicated to estimation of GAMMA-400 sensitivity to the narrow spectral lines due to DM annihilation or decay in the Galactic center (GC) for a quite generic DM candidate. Section \ref{sec:sn} analyzes the potential to discover/constrain ALPs by an observation of a nearby supernova explosion. Section \ref{sec:alp} is dedicated to estimation of GAMMA-400 sensitivity to ALPs by observations of bright gamma-ray pulsars. Section \ref{sec:misc} briefly discusses other targets: globular clusters, AGNs and others. And section \ref{sec:last} summarizes the obtained results.

In this section we briefly describe the expected GAMMA-400 performance based on our detailed simulations. In general, our telescope is capable for both photon and electron/positron detection (without a possibility to distinguish between electrons and positrons). The lower boundary of sensitivity range is $\approx$ 20 MeV and determined by a sophisticated combination of various factors like backgrounds, detector noises, Coulomb scattering of a pair produced in the converter-tracker etc. The upper boundary is determined mainly by statistical flux limitations and reaches $\sim$ 10 TeV for $e^\pm$ and $\sim$ 1 TeV for gammas. The telescope field of view determined analogously to that of Fermi-LAT is $\approx$ 1 sr. In general, our instrument is designed for the dedicated pointed deep observations of several interesting regions of the gamma-ray sky oppositely to the uniform survey strategy of Fermi-LAT.

\begin{figure}[h]
\centering 
\includegraphics[width=0.8\textwidth]{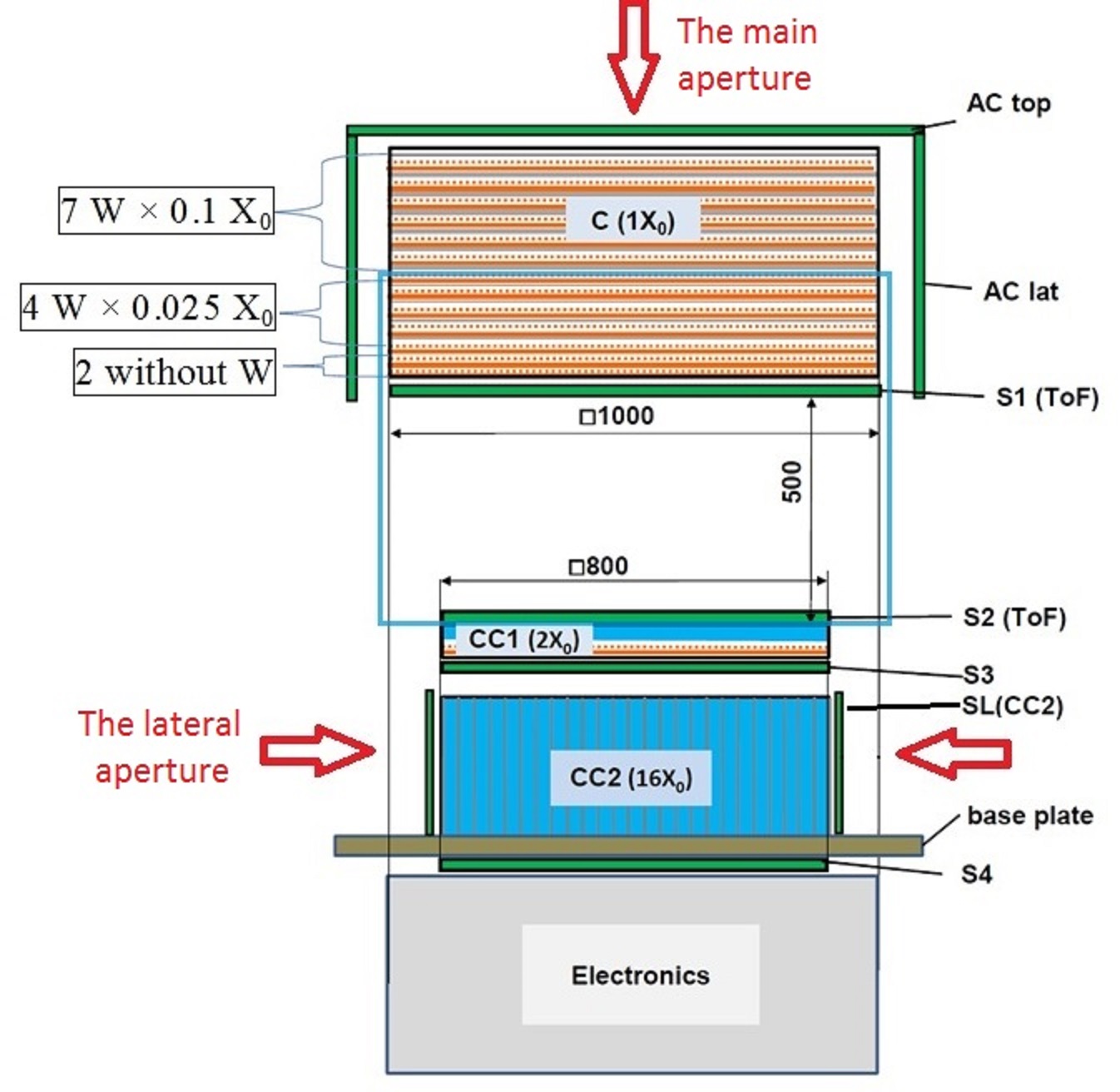}
\caption{\label{fig:sch} (COLOR ONLINE) The GAMMA-400 physical scheme. All sizes are in mm. ``AC'' denotes the anticoincidence system, ``C'' is the converter-tracker, ``ToF'' is the time-of-flight system and ``CC'' is the coordinate calorimeter. The big cyan rectangle encloses the part of the telescope used in the detection mode of low-energy photons by the thin layers of the converter-tracker referred below as the thin converter. More details see in section \ref{sec:i}.}
\end{figure}

\begin{figure}[H]
\centering  
\begin{subfigure}[t]{0.495\textwidth}
\includegraphics[width=\linewidth]{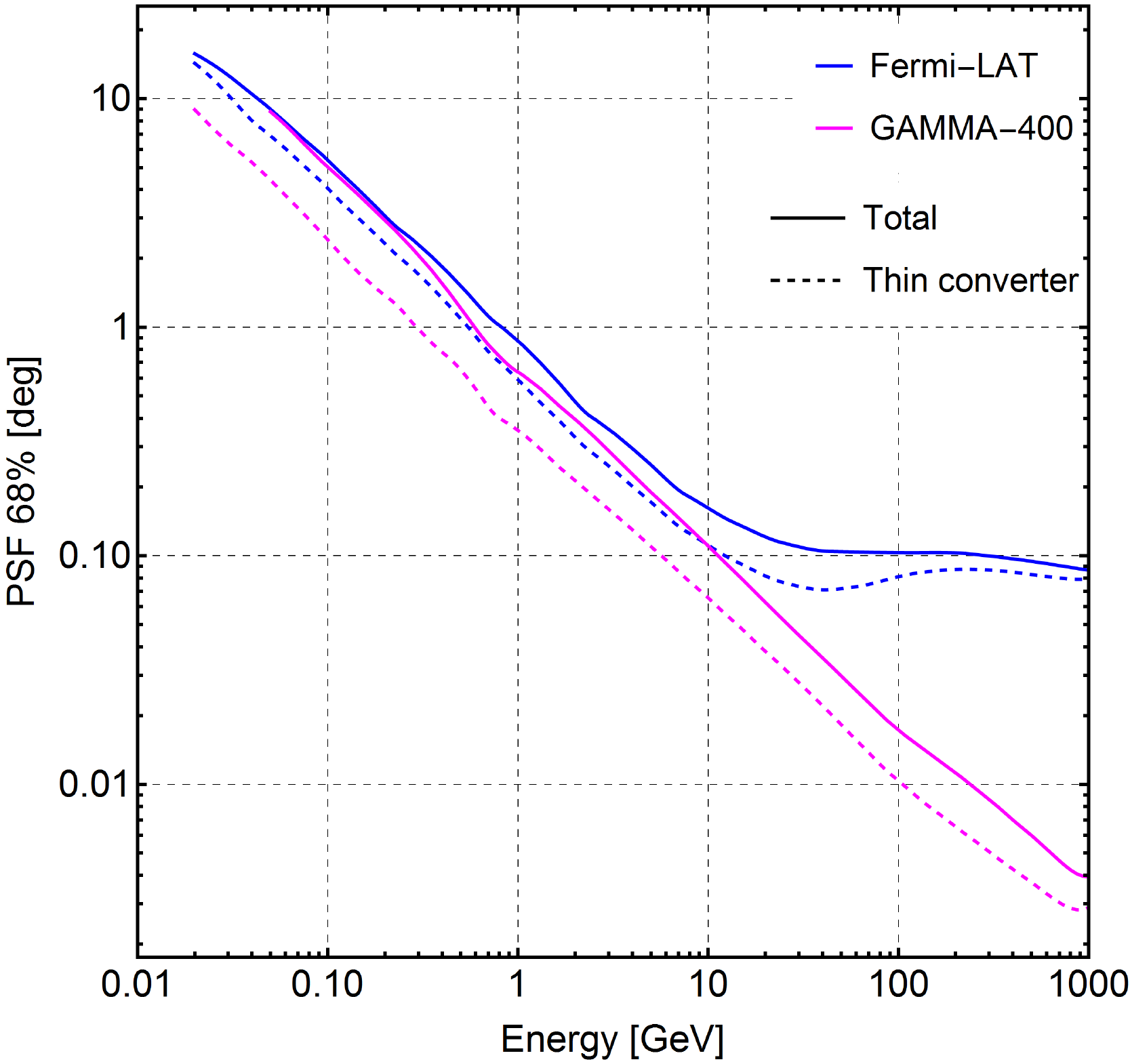}
\end{subfigure}
\begin{subfigure}[t]{0.495\textwidth}
\includegraphics[width=\linewidth]{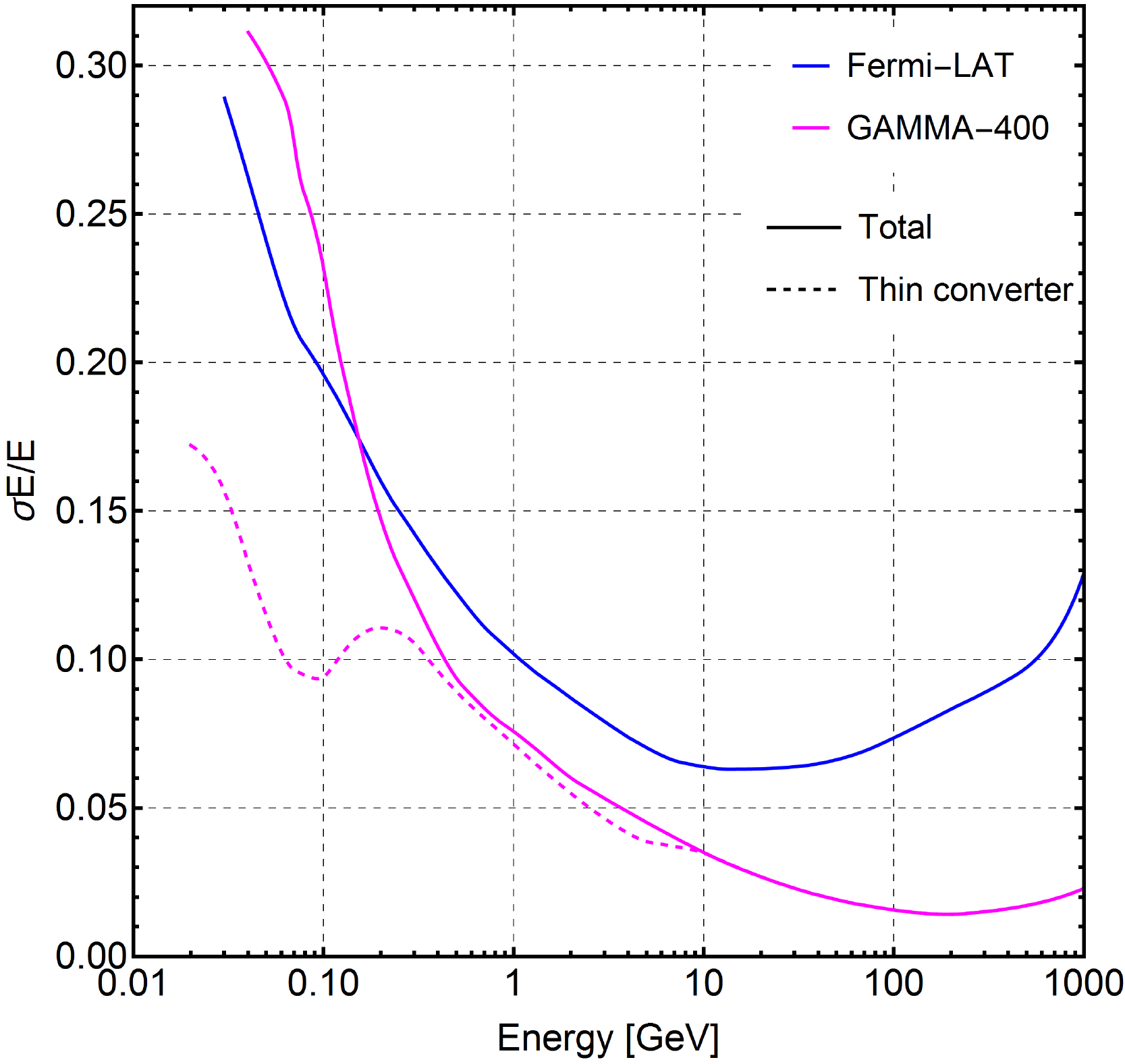}
\end{subfigure}
\vspace{0.35cm}
\begin{subfigure}[t]{0.495\textwidth}
\includegraphics[width=\linewidth]{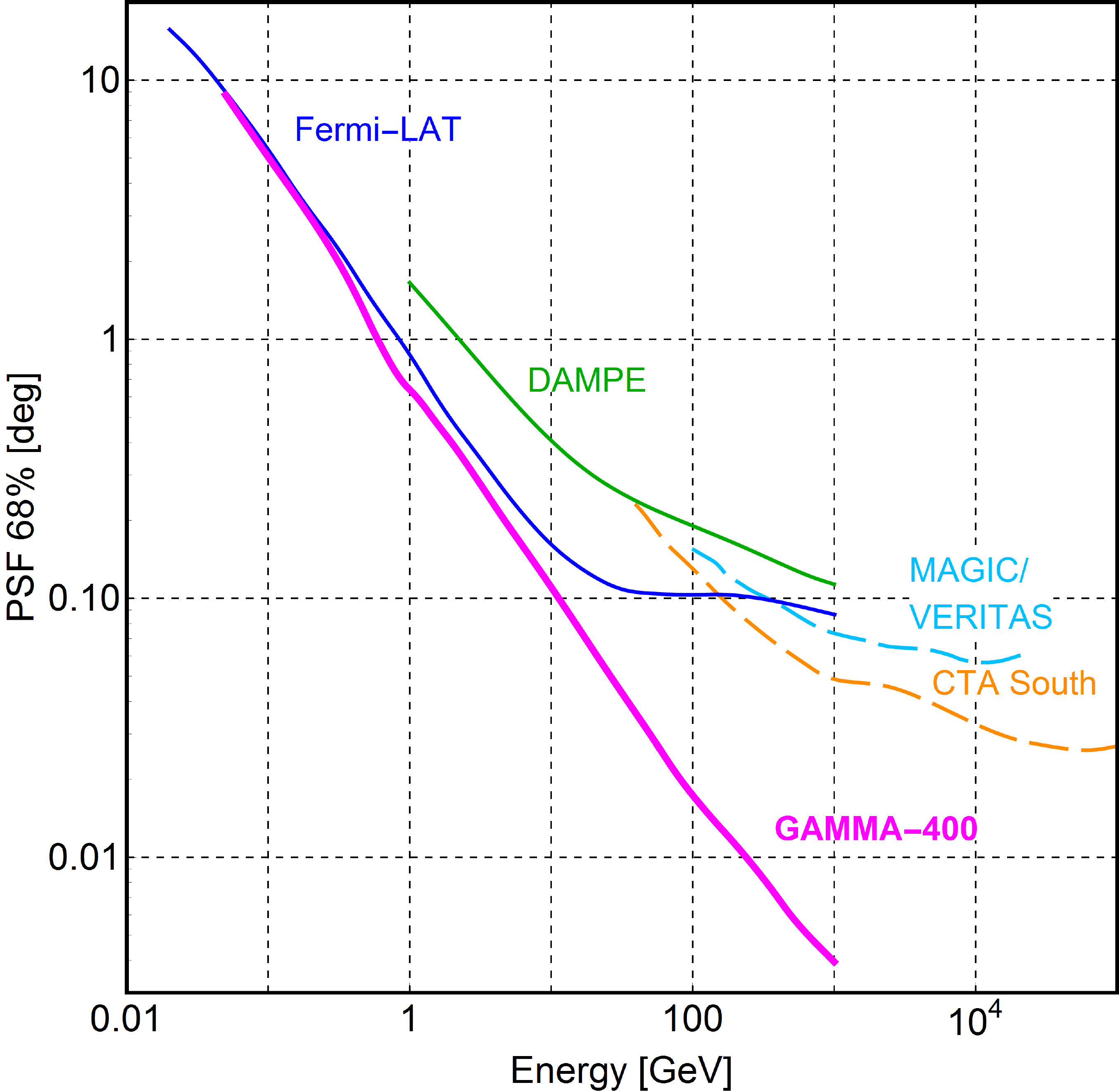}
\end{subfigure}
\begin{subfigure}[t]{0.495\textwidth}
\includegraphics[width=\linewidth]{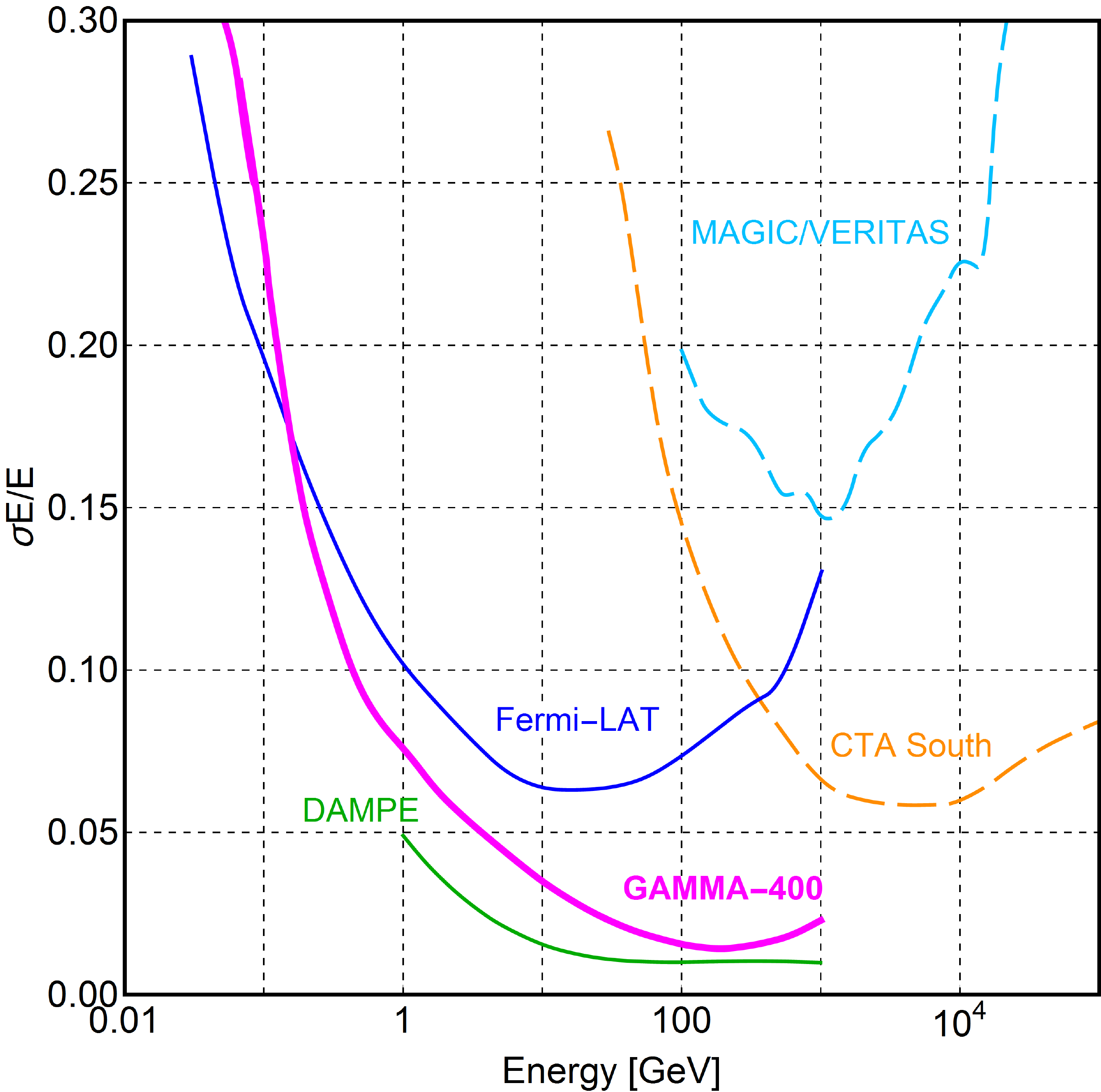}
\end{subfigure}
\vspace{0.35cm}
\begin{subfigure}[t]{0.495\textwidth}
\vspace{0pt}
\includegraphics[width=\linewidth]{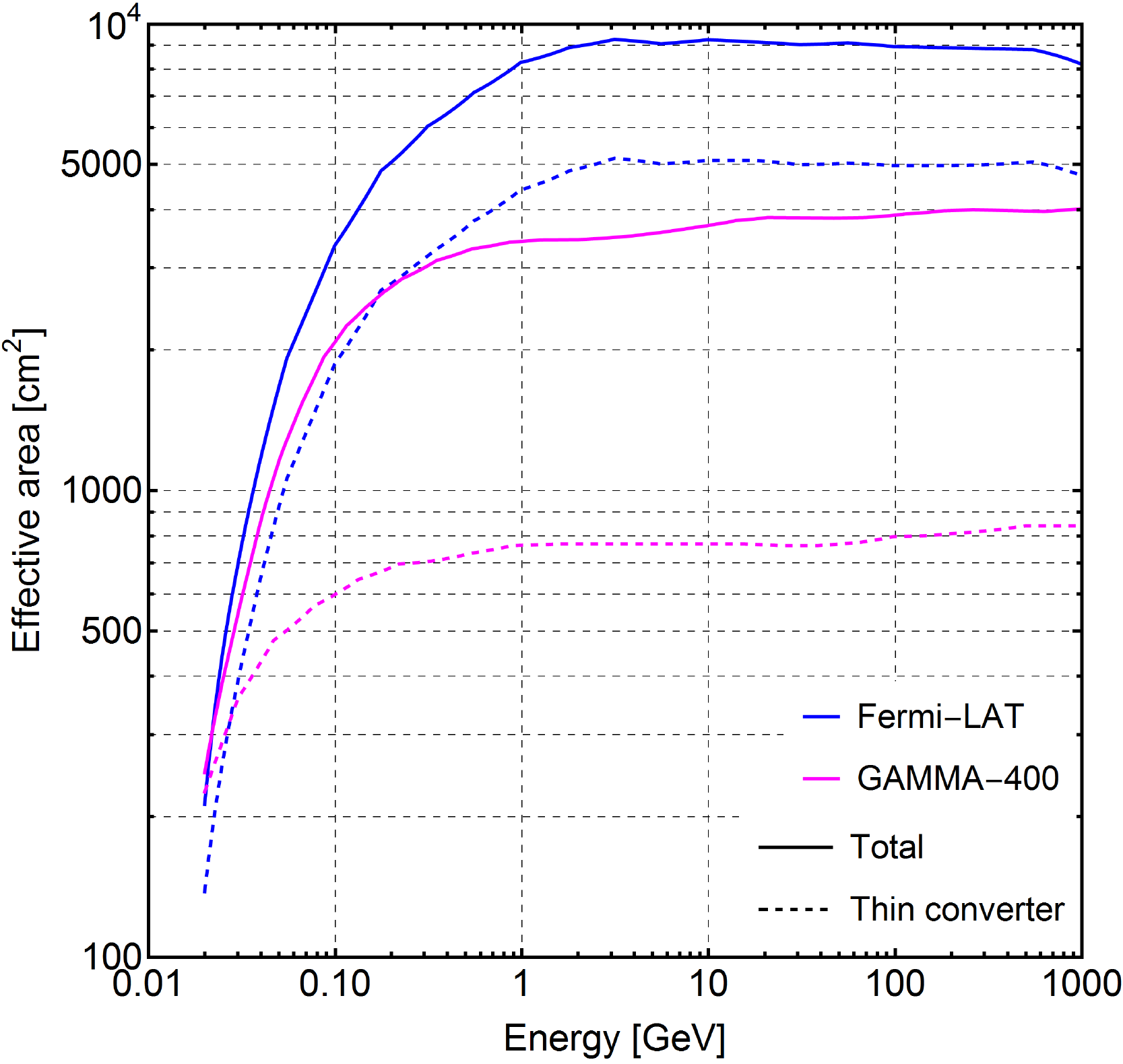}
\end{subfigure}
\begin{minipage}[t]{0.495\textwidth} \vspace{-0.5cm}
\caption{\label{fig:char} The energy dependence of the key characteristics of GAMMA-400 in comparison with other telescopes (performance data for Fermi-LAT was taken from \cite{Fermi}). \textit{\textbf{Top left:}} the angular resolution as the radius of point spread function (PSF) at the level of 68\% containment. \textit{\textbf{Top right:}} the energy resolution as the relative standard deviation of the measured photon energy distribution. \textit{\textbf{Middle left:}} GAMMA-400 and Fermi-LAT (whole converter-tracker) angular resolution in comparison with DAMPE \cite{2017ICRC...35..775D} and the ground-based Cherenkov telescopes MAGIC \cite{2016APh....72...76A}, VERITAS \cite{2015ICRC...34..771P} and CTA South \cite{CTA}. \textit{\textbf{Middle right:}} Same for the energy resolution. \textit{\textbf{Bottom left:}} the effective area for the cases of using the whole converter-tracker and only its ``thin'' part, which provides better performance at low energies. See more details in section \ref{sec:i}. }
\end{minipage}
\end{figure}

Figure \ref{fig:sch} shows the GAMMA-400 physical scheme, figure \ref{fig:char} demonstrates the dependence of all key characteristics on the energy inside the sensitivity range in comparison with other gamma-ray telescopes.   
The effective area, angular and energy resolutions are shown in comparison with Fermi-LAT for two cases - using the whole converter-tracker and only its ``thin'' part. The latter for Fermi-LAT is also often referred as Fermi-Front \cite{Fermi}. GAMMA-400 has essentially the same feature in the form of 4 thin tungsten layers of 0.025$X_0$ (radiation length units) and 2 layers without tungsten at the back of the converter-tracker (see figure \ref{fig:sch}). These thin layers provide significantly better angular and energy resolutions (see the dashed lines in figure \ref{fig:char}) due to reduced Coulomb scattering of the produced pair being tracked. Indeed the effective area of the thin converter is significantly lower than the total area. However at low energies it is compensated by large photon fluxes. In summary, we can see that the angular resolution of the thin converter of GAMMA-400 significantly exceeds that of Fermi-LAT at any energy. The whole converter of GAMMA-400 outperforms Fermi-LAT at energies above $\approx$ 10 GeV and reaches $\sim 0.01\degree$ at 100 GeV. The energy resolution superiority is even greater: it begins from $\approx$ 200 MeV (for the whole converter) and reaches $\approx$ 2\% at 100 GeV (the thin converter of Fermi-LAT does not differ much from the whole one by energy resolution \cite{Fermi}). The comparison with DAMPE and ground-based telescopes is also shown. 

All characteristics described above refer to the classical mode of photon detection, when the photon enters the converter without interaction in the anticoincidence system and produces a pair, which releases its energy in the calorimeter. Such a mode can be called the main aperture detection. However, in principle one can detect photons entering the calorimeter directly from its lateral surfaces (see figure \ref{fig:sch}). For this purpose our telescope will have the lateral detectors of calorimeter (SL(CC2)). We call this mode the lateral aperture detection. Indeed it has a photon direction reconstruction worse than that for the main aperture. However, it may be still useful for some tasks like a narrow line search in a large-scale diffuse background, detection of GRBs and other transients etc. For this reason we conduct a dedicated modeling of the lateral aperture \cite{2020PAN....82..845A}. The preliminary modeling results showed a good performance of the lateral aperture: the lower detection threshold is $\sim$ 10 MeV, the effective area is up to $\approx$ 2000 cm$^2$ for each side and a large total field of view is about 2--3 sr. 

Also we would like to clarify in this section one important aspect: our telescope is under designing stage yet and, hence, several differences in its description exist in the papers published in different years. The telescope version described in this paper is already quite stable. At the same time, as can be seen in other papers (e.g. \cite{2019JPhCS1181a2041T}), a larger version of GAMMA-400 is also under consideration. This version has a bigger calorimeter in all three dimensions ($1 \times 1 ~\mbox{m}^2$ and $\approx 25X_0$), which provides the effective area up to 5500 cm$^2$ and the energy resolution $\approx$ 1\% at high energies. But such an extended version has the mass of $\approx$ 4 t instead of $\approx$ 2 t for the base version described in the previous paragraphs. Currently we do not have a certainty yet which launch vehicle will be provided for GAMMA-400. Hence we consider essentially both versions: the base one as rather guaranteed and the extended one as an optimistic case, when a heavy enough launch vehicle and additional funds will be provided. In this paper we mainly describe and build predictions for the base version.

\section{GAMMA-400 sensitivity to the spectral lines due to DM annihilation or decay in the Galactic center}
\label{sec:l}
Historically GAMMA-400 was designed to have the thick calorimeter ($\gtrsim 20X_0$) in order to provide an excellent energy resolution and, hence, the sensitivity to narrow spectral lines. The latter can be generated by a big variety of DM candidates through annihilation or decay. Besides traditional WIMPs annihilating directly to photons, this variety includes: Kaluza-Klein particles \cite{2012JCAP...03..020B}, the hidden sector DM \cite{2019PhRvD.100l3007F}, gravitino \cite{2014JCAP...10..023A}, pseudo-Goldstone DM \cite{2019PhRvD.100c5023C}, secluded DM \cite{2020JHEP...07..148Y} and others. Thus the line search represents a very generic and powerful DM search method - it is not tuned to a specific candidate, but rather tests many different candidates simultaneously. 

According to the preliminary observational program GAMMA-400 will conduct the deep continuous observations of the GC during 2--4 years. The GC is known to be the best target in the sky for the search of lines from annihilating DM - notwithstanding bright backgrounds - since it has the biggest J-factor among all the targets. There were published already several papers on the line constraints by Fermi-LAT observations \cite{2014JCAP...10..023A,2015PhRvD..91l2002A}. We followed the standard methodology used in these papers in order to estimate the line sensitivity of GAMMA-400 alone and together with Fermi-LAT. We calculated the sensitivity to both annihilating and decaying DM, i.e. the limiting velocity-averaged diphoton annihilation cross section $\langle \sigma v \rangle_{\chi\chi\rightarrow\gamma\gamma}$ for the range of DM particle masses $m_\chi$ from 1 to 450 GeV and the limiting DM lifetime $\tau_{\chi\rightarrow\gamma\nu}$ for the mass range (1--900) GeV (analogically to \cite{2015PhRvD..91l2002A}). We calculated the limiting cross sections and lifetimes for $\sim$ 10 discrete mass values in those ranges and then interpolated between the values. To estimate the line sensitivity we largely followed the methodology described in \cite{2014JCAP...10..023A}. Thus we constructed the following likelihood function, which is essentially the probability density for $\mu_{sig}\left({\langle \sigma v \rangle_{\gamma\gamma} \atop \tau_{\gamma\nu}}, m_\chi\right)$ - the average number of photons in the bin from DM line in the case of observation of $n$ photons in total and expectation of $\mu_b$ background photons (in the bin - bin index is omitted, bins are independent):
\begin{equation}
	L\left(\mu_{sig}\left({\langle \sigma v \rangle_{\gamma\gamma} \atop \tau_{\gamma\nu}}, m_\chi\right),n_s|n,\mu_b \right) \propto P_s(n_s,\mu_b) \times P(\mu|n) = \frac{1}{\sqrt{2\pi}\sigma_s}\exp\left(-\frac{n_s^2}{2\sigma_s^2}\right) \times \frac{\mu^n}{n!}e^{-\mu},
	\label{eq:L}
\end{equation}
where $P_s(n_s,\mu_b)$ reflects the systematic flux uncertainty and modeled by the normal distribution, $P(\mu|n)$ is the standard Poisson distribution of the photon count, in a general case $n=n_b+n_{sig}+n_s$, $\mu=\mu_b+\mu_{sig}+n_s$, $n_s$ is the photon number systematic offset due to CR contamination etc. (nuisance parameter), $n_b~(\mu_b)$ is the measured (average) number of background photons, $\sigma_s$ is the standard deviation of the systematic offset and taken to be $\sigma_s = \delta f_s \times \mu_b$ analogically to \cite{2014JCAP...10..023A}. The fractional systematic uncertainty $\delta f_s$ is assumed to be the same for both telescopes and equal to 0.015 according to \cite{2015PhRvD..91l2002A} except the case of the search of decaying DM by Fermi-LAT, when it collects the signal from the whole sky with larger $\delta f_s = 0.018$. To estimate the median sensitivity we put $n=\mu_b$. Let us expand in details $\mu_b$ and $\mu_{sig}$:
\begin{equation}
	\mu_b(m_\chi)=\int\limits_{m_\chi-k(m_\chi)\sigma_E(m_\chi)}^{m_\chi+k(m_\chi)\sigma_E(m_\chi)} dE' \int dE f_b(E)\frac{1}{\sqrt{2\pi}\sigma_E(E)}\exp\left(-\frac{(E-E')^2}{2\sigma_E^2(E)}\right)\varepsilon(E),
	\label{eq:mb}
\end{equation}
where $f_b(E)$ is the background spectrum and $\varepsilon(E)=\int A_{eff}(E,t)dt$ is the exposure with $A_{eff}(E,t)$ being the telescope effective area, which is observing the target at an arbitrary moment. This number of photons is calculated (for the annihilation case) inside the energy bin $m_\chi \pm k(m_\chi)\sigma_E(m_\chi)$, where $k(m_\chi)$ is the energy bin half-width in the units of $\sigma_E(E=m_\chi)$, i.e. the RMS of energy dispersion of the telescope or essentially its energy resolution (shown in figure \ref{fig:char}). We tuned $k(m_\chi)$ to maximize the sensitivity being estimated for each DM particle mass (see also below). However the sensitivity does not depend strongly on $k(m_\chi)$. The second integral over $E$ in \eqref{eq:mb} technically goes over all energies. However practically we integrated over the range $E' \pm 5\sigma_E(E')$, which is absolutely enough. The background spectrum $f_b(E)$ inside the chosen region of interest (ROI, will be described below) is assumed to be a priori known: we extracted it from the Fermi-LAT background model map \cite{Fermi-map} using the Aladin sky atlas \cite{2000A&AS..143...33B}. Here we included only the Galactic diffuse component, since the contribution of point sources and the isotropic background in the GC region is small ($\lesssim$ 10\%, see e.g. figure 1 in \cite{2017ApJ...840...43A}) and was checked to be negligible. The exponent function in \eqref{eq:mb} models the energy dispersion of telescopes.

As the first step we reproduced the Fermi-LAT line limits published in \cite{2015PhRvD..91l2002A} based on 6 years of observations in order to verify a correctness of our procedure. For this calculation we needed in the Fermi-LAT exposure $\varepsilon(E)$ and took it from \cite{2020ApJS..247...33A}. Although \cite{2020ApJS..247...33A} provides the exposure value only at 1 GeV for 8 years of operation, we extrapolated it to an arbitrary energy and operation time by simple rescaling through the effective area: $\varepsilon_F(E,T) \approx \varepsilon_F(1~\mbox{GeV},8~\mbox{years}) \frac{A_{eff}(E)}{A_{eff}(1~\mbox{GeV})} \frac{T}{8~\mbox{years}}$.	

Now let us discuss calculation of the expected DM line signal $\mu_{sig}\left({\langle \sigma v \rangle_{\gamma\gamma} \atop \tau_{\gamma\nu}}, m_\chi\right)$ and the ROI for the line search. The DM signal (as the photon flux density) is calculated through the well-known equations (e.g. \cite{2015PhRvD..91l2002A}):
\begin{gather}
	\frac{d\Phi_\gamma^{ann}}{dE} = \frac{1}{8\pi} \frac{\langle \sigma v \rangle_{\gamma\gamma}}{m_\chi^2} \left(\frac{dN_\gamma}{dE}\right)_{\gamma\gamma} J(\Omega_{ROI}), ~\frac{d\Phi_\gamma^{dec}}{dE} = \frac{1}{4\pi\tau_{\gamma\nu} m_\chi} \left(\frac{dN_\gamma}{dE}\right)_{\gamma\nu} D(\Omega_{ROI}), \\ 
	J(\Omega_{ROI})||D(\Omega_{ROI}) = \int\limits_{l.o.s.}dl\int\limits_{ROI}d\Omega~ \rho^{2||1}(r=\sqrt{r_\odot^2+l^2-2r_\odot l \cos\theta}),
\label{eq:J}
\end{gather}
where $\rho(r)$ is the DM density dependence on the distance from the GC and $\left(\frac{dN_\gamma}{dE}\right)_{\gamma\gamma/\gamma\nu}$ is the photon spectrum from one annihilation/decay. One annihilation yields two photons with the energies equal to the rest energy of DM particle. One decay may yield various pairs: two photons, photon and neutrino, photon and Z-boson etc. We followed the choice of \cite{2015PhRvD..91l2002A}, where DM decays to the photon and neutrino with energies equal to the half of DM rest energy. The majority of papers dedicated to the line searches (e.g. \cite{2014JCAP...10..023A,2015PhRvD..91l2002A,2016JCAP...02..026A,2017PhRvD..95f3531L}) assumes the spectrum to be essentially the delta-function: $\left(\frac{dN_\gamma}{dE}\right)_{\gamma\gamma} = 2\delta(E-m_\chi)$ and $\left(\frac{dN_\gamma}{dE}\right)_{\gamma\nu} = \delta\left(E-\frac{m_\chi}{2}\right)$. This is indeed true for virtually still DM particles. However any DM halo has a velocity dispersion, which inevitably leads to the Doppler broadening of the line. This is especially relevant for large halos like galaxy clusters considered in e.g. \cite{2016JCAP...02..026A}. For the case of our Galaxy we expect the relative Doppler broadening to be $\sim 10^{-3}$. However with GAMMA-400 we deal with energy resolutions down to $\sim 10^{-2}$ level, which is not drastically larger than the line width mentioned. For this reason we decided to check precisely whether we can safely ignore the finite line width. For that we modeled the line shape as follows (the annihilation case, decay line width is the same):
\begin{equation}
	\left(\frac{dN_\gamma}{dE}\right)_{\gamma\gamma} \approx \frac{1700}{m_\chi}\exp\left(-\frac{(E-m_\chi)^2}{2\sigma_l^2(m_\chi)}\right),~ \sigma_l(m_\chi) = \frac{\langle v_{l.o.s.}\rangle m_\chi}{\sqrt{2\ln2}c},
	\label{eq:l}
\end{equation}
where $\langle v_{l.o.s.}\rangle \approx 200$ km/s \cite{Bertone-book} and the normalization coefficient was obtained from the simple condition $\int\left(\frac{dN_\gamma}{dE}\right)_{\gamma\gamma}dE = 2$. Then we tested the difference in the number of photons from line calculated with the simple delta-like line shape and the Gaussian one \eqref{eq:l} inside the (narrowest) energy bin. We obtained a negligible difference and then confidently proceeded to work with effectively monochromatic photons from the line.

As the DM density profile we chose a very typical one - Einasto profile with parameters from \cite{2015PhRvD..91l2002A}:
\begin{equation}
	\rho (r) = \rho_s \exp\left(-\frac{2}{\alpha}\left(\left(\frac{r}{r_s}\right)^{\alpha}-1\right)\right),
	\label{eq:rho}
\end{equation}
where $\rho_s$ = 0.081 GeV/cm$^3$, $r_s$ = 20 kpc and $\alpha$ = 0.17. This yields $\rho(r=8.5$~kpc) = 0.4 GeV/cm$^3$. We did not include substructure boost due to its smallness in the central region of the halo \cite{2010PhRvD..81d3532K}. Meanwhile, one basic sanity check for DM density profile is to test whether it yields the correct whole halo mass, which is quite well known for our Galaxy to be $M_{200} \approx 1.2 \cdot 10^{12} M_\odot$ \cite{2019MNRAS.484.5453C}. To calculate the whole halo mass for our chosen profile \eqref{eq:rho} we used the following equation for the Einasto enclosed mass from \cite{2005MNRAS.358.1325C,2005MNRAS.362...95M}:
\begin{equation}
	M(r<R,r_s,\rho_s,\alpha) = \int\limits_0^R 4\pi r^2 \rho(r) dr = \frac{4\pi \rho_s r_s^3}{\alpha} \left(\frac{2}{\alpha}\right)^{-\frac{3}{\alpha}} e^{2/\alpha} \left(\Gamma\left(\frac{3}{\alpha}\right) - \Gamma\left(\frac{3}{\alpha},\frac{2}{\alpha}\left(\frac{R}{r_s}\right)^\alpha\right)\right),
	\label{eq:ein}
\end{equation}
where $\Gamma(z) = \int\limits_0^\infty x^{z-1}e^{-x}dx$ is the usual and $\Gamma(z,a) = \int\limits_a^\infty x^{z-1}e^{-x}dx$ is the upper incomplete gamma functions. The equation above yielded $M(R \approx 10r_s) \approx 1.4 \cdot 10^{12} M_\odot$, which is in a reasonable agreement with the value cited above from \cite{2019MNRAS.484.5453C} (neglecting by substructures and baryons) taking into account their uncertainties.

For the ROI for the annihilation case we naturally chose the one from \cite{2015PhRvD..91l2002A} in their case of Einasto profile, i.e. the disk with radius $\theta_{max} = 16\degree$ around the GC. We essentially copied the density profile and ROI from \cite{2015PhRvD..91l2002A} due to two reasons: at first, as was already mentioned, we wanted to check whether our algorithm correctly reproduces the Fermi-LAT constraints. And secondly we would like to make a pure comparison of GAMMA-400 line sensitivity with that of Fermi-LAT under the similar circumstances. For the decay case the situation is more tricky. The decay signal is less concentrated towards GC in the sky, and it is more efficient to collect such signal from a large part of the sky. Thus in \cite{2015PhRvD..91l2002A} the whole sky (i.e. $\theta_{max} = 180\degree$) was used for this purpose. However as was mentioned above, GAMMA-400 is intended to deeply observe only several selected areas of interest in the sky with the biggest exposure dedicated to the GC. Hence in order to estimate our sensitivity to decaying DM we naturally chose the same ROI as for the annihilation case. For this ROI the chosen profile \eqref{eq:rho} yields for \eqref{eq:J} $J(\theta_{max} = 16\degree) \approx 1.1 \cdot 10^{23}~\mbox{GeV}^2/\mbox{cm}^5$ and $D(\theta_{max} = 16\degree) \approx 2.6 \cdot 10^{22}~\mbox{GeV}/\mbox{cm}^2$. 

At this point we could determine $\mu_{sig}\left({\langle \sigma v \rangle_{\gamma\gamma} \atop \tau_{\gamma\nu}}, m_\chi\right)$ for \eqref{eq:L}:
\begin{equation}
\begin{aligned}
&\mu_{sig}(\langle \sigma v \rangle_{\gamma\gamma}, m_\chi) = \int\limits_{m_\chi-k(m_\chi)\sigma_E(m_\chi)}^{m_\chi+k(m_\chi)\sigma_E(m_\chi)} dE' \int dE \frac{d\Phi_\gamma^{ann}}{dE}(E)\frac{1}{\sqrt{2\pi}\sigma_E(E)}\exp\left(-\frac{(E-E')^2}{2\sigma_E^2(E)}\right)\varepsilon(E) \\  
& ~~~~~~~~~~~~~~~~~~~~~~~~~~~~~~~~~~~~~~~~~~~~~~~~~~~~~~	= \frac{\langle \sigma v \rangle_{\gamma\gamma}}{4\pi m_\chi^2} J(\Omega_{ROI}) \mbox{erf}\left(\frac{k(m_\chi)}{\sqrt{2}}\right) \varepsilon(m_\chi), \\ 
&\mu_{sig}(\tau_{\gamma\nu}, m_\chi) =	\int\limits_{\frac{m_\chi}{2}-k(\frac{m_\chi}{2})\sigma_E(\frac{m_\chi}{2})}^{\frac{m_\chi}{2}+k(\frac{m_\chi}{2})\sigma_E(\frac{m_\chi}{2})} dE' \int dE \frac{d\Phi_\gamma^{dec}}{dE}(E)\frac{1}{\sqrt{2\pi}\sigma_E(E)}\exp\left(-\frac{(E-E')^2}{2\sigma_E^2(E)}\right)\varepsilon(E) \\
& ~~~~~~~~~~~~~~~~~~~~~~~~~~~~~~~~~~~~~~~~~~~~~~~~~~~~~~ = \frac{1}{4\pi\tau_{\gamma\nu}m_\chi} D(\Omega_{ROI}) \mbox{erf}\left(\frac{k(m_\chi/2)}{\sqrt{2}}\right) \varepsilon(m_\chi/2), 
	\label{eq:ms}
\end{aligned}
\end{equation}
where $\mbox{erf}(x) = \frac{2}{\sqrt{\pi}} \int\limits_0^x e^{-t^2}dt$ is the error function, and we took into account the approximation of the line shape by delta-function discussed above. Finally we determined the following condition for the median sensitivity at 95\% confidence level:
\begin{equation}
\begin{aligned}
	\int\limits_0^{\langle \sigma v \rangle_{\gamma\gamma}^{lim}} d\langle \sigma v \rangle_{\gamma\gamma} \int dn_s L(\mu_{sig}(\langle \sigma v \rangle_{\gamma\gamma},m_\chi),n_s|n=\mu_b,\mu_b) &= 0.95, \\
	\int\limits_{\tau_{\gamma\nu}^{lim}}^\infty d\tau_{\gamma\nu} \int dn_s L(\mu_{sig}(\tau_{\gamma\nu},m_\chi),n_s|n=\mu_b,\mu_b) &= 0.95,
	\label{eq:0.95}
\end{aligned}
\end{equation}
i.e. we marginalized the likelihood over the nuisance parameter $n_s$. This is also often called the Bayesian approach (with the flat prior, see e.g. \cite{2020JCAP...02..012H}). Then we substituted \eqref{eq:L},\eqref{eq:mb},\eqref{eq:ms} into \eqref{eq:0.95} and found numerically the limiting values of $\langle \sigma v \rangle_{\gamma\gamma}^{lim}$ and $\tau_{\gamma\nu}^{lim}$ for each $m_\chi$. For masses $m_\chi \lesssim 100$ GeV the photon counts are large, and we used the well-known approximation of the Poisson distribution $P(\mu|n) = \frac{\mu^n}{n!}e^{-\mu} \approx \frac{1}{\sqrt{2\pi\mu}}\exp\left(-\frac{(n-\mu)^2}{2\mu}\right)$ for large $\mu$.

\begin{figure}[h]
\centering 
\includegraphics[width=1\textwidth]{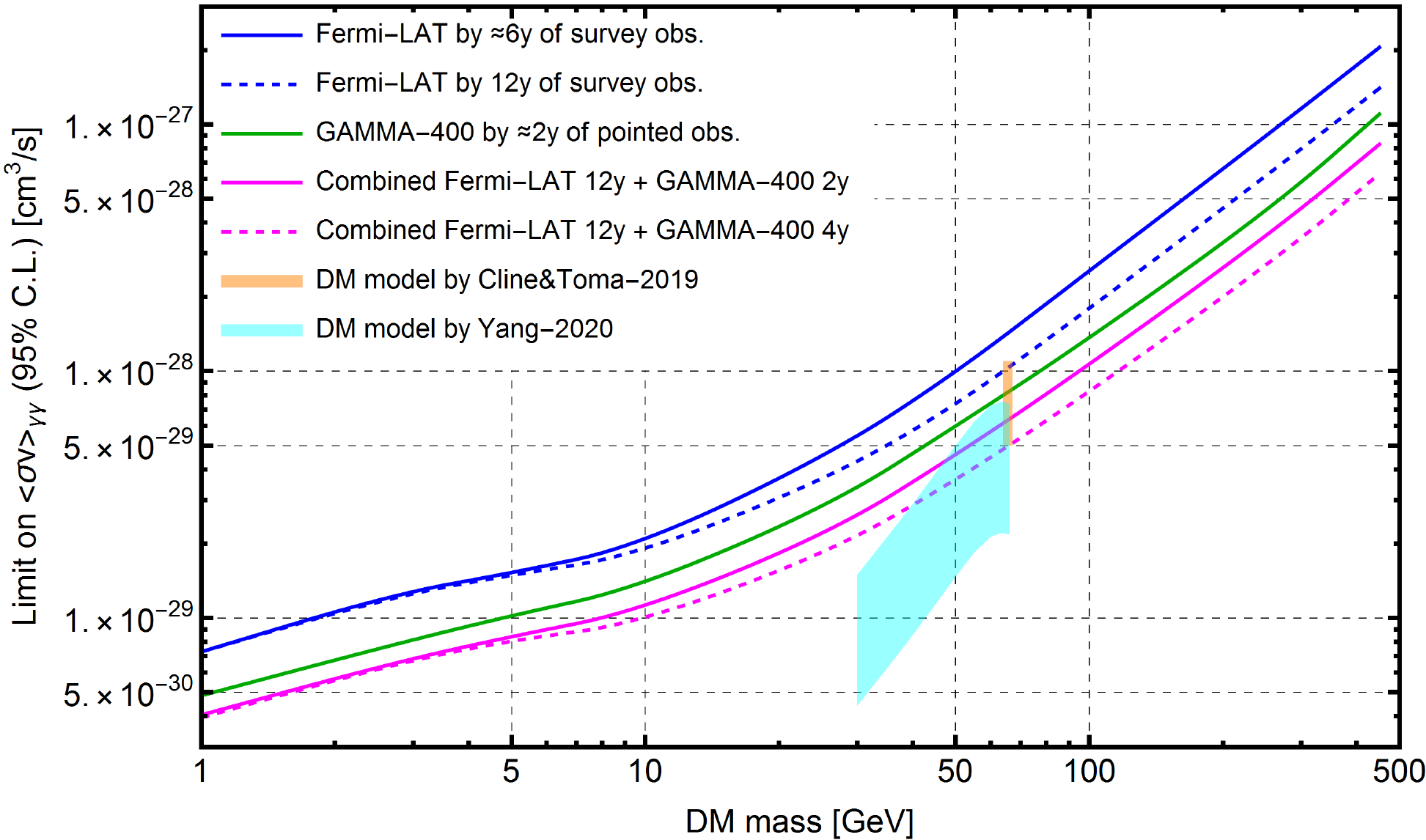}
\caption{\label{fig:sv} GAMMA-400 median sensitivity to the diphoton annihilation cross section in comparison with that of Fermi-LAT (alone and combined) for the case of Einasto DM density profile and the ROI radius of 16$\degree$ around the GC. The predictions of two specific DM models \cite{2019PhRvD.100c5023C,2020JHEP...07..148Y} are also shown, which could be directly tested. More details see in section \ref{sec:l}.}
\end{figure}

\begin{figure}[h]
\centering 
\includegraphics[width=1\textwidth]{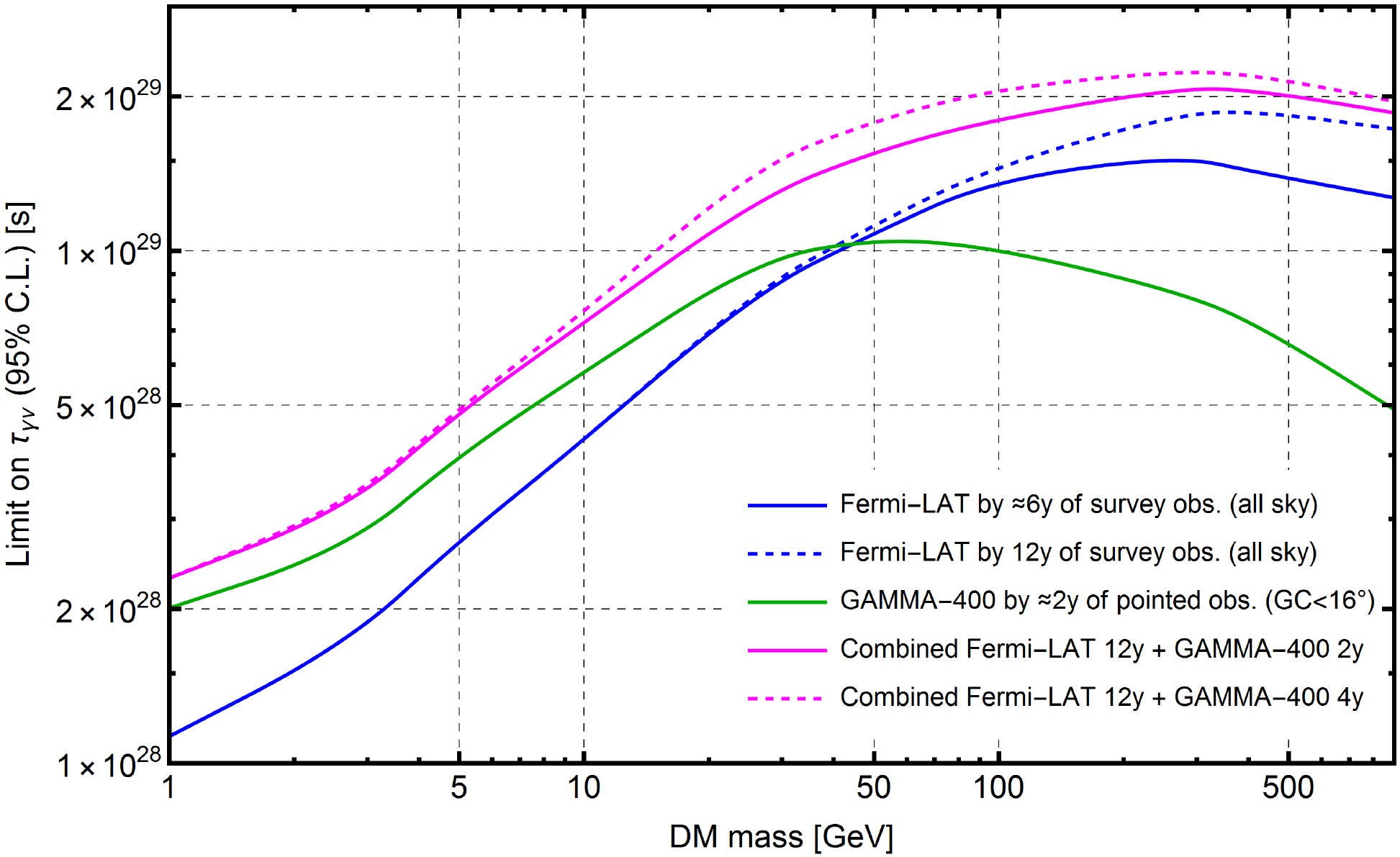}
\caption{\label{fig:tau} GAMMA-400 median sensitivity to the DM lifetime in comparison with that of Fermi-LAT (alone and combined) for the case of Einasto DM density profile. GAMMA-400 collects the signal from the disk with 16$\degree$ radius around the GC, Fermi-LAT - from all sky. More details see in section \ref{sec:l}.}
\end{figure}

As we already mentioned, firstly we obtained the sensitivity (i.e. $\langle \sigma v \rangle_{\gamma\gamma}^{lim}(m_\chi)$ and $\tau_{\gamma\nu}^{lim}(m_\chi)$) for Fermi-LAT by 5.8 years of observations in order to compare our result with \cite{2015PhRvD..91l2002A} to be sure in the correctness of our algorithm. Our obtained sensitivity is shown by the blue lines in figures \ref{fig:sv}--\ref{fig:tau}. It matches the sensitivity obtained in \cite{2015PhRvD..91l2002A} with the accuracy not worse than $\approx$ 20\% for all the masses. In our view this is a very good concordance for our relatively simplistic procedure.\footnote{A little detail worthy to mention is that \cite{2015PhRvD..91l2002A} used NFW density profile for the case of decaying DM. However D-factors \eqref{eq:J} are almost exactly the same for NFW and Einasto profiles (for $\Omega_{ROI} = 4\pi$ sr). Hence our sensitivity with Einasto profile corresponds directly to that with NFW in \cite{2015PhRvD..91l2002A} (fig. 9 there).} This assured us and we proceeded to the next step: prediction of Fermi-LAT sensitivity for its current data set, i.e. $\approx$ 12 years of observations. This prediction is shown by the blue dashed lines in figures \ref{fig:sv}--\ref{fig:tau}. We may conclude the following: at low DM masses $m_\chi \lesssim 10$ GeV Fermi-LAT has clearly reached a saturation, i.e. an increase of exposure does not increase the sensitivity at all (background-dominated regime). In other words, the telescope's energy resolution sets a fundamental limit to the line sensitivity, which can not be overcome by an exposure increase. At higher DM masses the signal to background ratio is larger, and the sensitivity still increases with exposure, however not so much. 

Then we calculated GAMMA-400 sensitivity for the annihilation case for the pointed (targeted) observations of the GC during $\approx$ 2 years, which is the minimal anticipated exposure time according to the preliminary GAMMA-400 observational program. As we already mentioned, we optimized the energy bin half-width $k(m_\chi)$ to maximize the sensitivity. We obtained that the optimal half-width monotonically increases with DM mass from $k(m_\chi = 1~\mbox{GeV}) = 0.35$ to $k(m_\chi = 450~\mbox{GeV}) = 1.65$. The calculated GAMMA-400 sensitivity is shown by the green line in figure \ref{fig:sv}. We see that it exceeds the sensitivity of Fermi-LAT even in the case of 12 years of its observations for all DM masses. The sensitivity gain is 1.2--1.5 times depending on DM mass. Such sensitivity is achievable by GAMMA-400 thanks to both a good energy resolution and the pointed observation mode. Thus the latter provides the exposure $\varepsilon_G(E=100~\mbox{GeV},T=2~\mbox{years}) \approx 2 \cdot 10^{11}~\mbox{cm}^2$s. For comparison, Fermi-LAT during 12 years has accumulated $\varepsilon_F(E=100~\mbox{GeV},T=12~\mbox{years}) \approx 6 \cdot 10^{11}~\mbox{cm}^2$s.

Then we naturally assumed that the future GAMMA-400 data can be jointly analyzed with the already existing Fermi-LAT data and estimated the combined sensitivity of both telescopes. For this we constructed the joint likelihood function as the product of likelihood functions for each telescope. Then our median sensitivity condition \eqref{eq:0.95} transforms to (up to a normalization factor):
\begin{equation}
\begin{aligned}
	\int\limits_0^{\langle \sigma v \rangle_{\gamma\gamma}^{lim}} d\langle \sigma v \rangle_{\gamma\gamma} \int dn_{sF} \int dn_{sG} L_F(\mu_{sig,F}(\langle \sigma v \rangle_{\gamma\gamma},m_\chi),n_{sF}|n=\mu_{bF},\mu_{bF}) \times \\ \times L_G(\mu_{sig,G}(\langle \sigma v \rangle_{\gamma\gamma},m_\chi),n_{sG}|n=\mu_{bG},\mu_{bG}) = 0.95,
	\label{eq:0.95j}
\end{aligned}
\end{equation}
where index ``F'' denotes everything for Fermi-LAT, ``G'' - for GAMMA-400. The combined sensitivity of Fermi-LAT by 12 years of survey observations and GAMMA-400 by 2 years of the pointed observations of the GC is shown by the magenta line in figure \ref{fig:sv}. As can be seen, such a joint analysis increases the sensitivity even further: by 1.6--1.8 times depending on DM mass with respect to Fermi-LAT by 12 years limits. Also we naturally expect that the combined limits will have much smaller uncertainty than the limits from a single telescope, although we did not evaluate the limit uncertainties here quantitatively. In the best case scenario our telescope can be operating during up to 10 years. In such case, in principle it can observe the GC during $\approx$ 4 years. We estimated the combined sensitivity for this case - it is shown by the dashed magenta line in figure \ref{fig:sv} and considered to be an optimistic scenario providing the sensitivity gain by 1.8--2.3 times. And in the meanwhile, we also estimated the sensitivity of the possible extended (4 t) version of GAMMA-400 mentioned in section \ref{sec:i}. The joint sensitivity of Fermi-LAT and such version of GAMMA-400 (4 years of exposure) will exceed Fermi-LAT alone by even much more - 1.8--3.5 times! 

Figure \ref{fig:sv} also contains the predictions of two specific DM models \cite{2019PhRvD.100c5023C,2020JHEP...07..148Y}. In \cite{2019PhRvD.100c5023C} the recent pseudo-Goldstone boson DM model was presented with $m_\chi = (64-67)$ GeV, which can explain the GC gamma-ray excess and the antiproton excess in CRs by the annihilation mainly into $b\bar{b}$. Also it can solve certain anomalies observed in LEP and CMS experiments. And at the same time this model predicts some non-zero annihilation branching ratio to photons, which implies the narrow line with the parameters marked by the orange region in figure \ref{fig:sv}. We see that Fermi-LAT alone can not reach this region to test. But together with GAMMA-400 the whole orange region can be tested! Similarly the cyan region shows the predictions of another model \cite{2020JHEP...07..148Y}, where DM has the form of secluded (vector) particles, which firstly annihilate into mediators with subsequent decay of the latter into SM particles including photons. This model also explains the GC gamma-ray excess and predicts the narrow line. However for this model the prediction is not so well-localized in the plane $\langle \sigma v \rangle_{\gamma\gamma} - m_\chi$ - we see that Fermi-LAT together with GAMMA-400 will be able to probe just a top part of the cyan region. Just to clarify we notice that indeed this vector DM does not annihilate to gammas directly. The cyan region in the figure reflects the effectively expected parameter values, if one would pretend diphoton annihilation to treat this candidate on the same parameter plane. Speaking more generally about other possible DM candidates including traditional WIMPs, theoretically we can expect to find them everywhere on the parameter plane below the level $\sim 0.1\langle \sigma v \rangle_{thermal} \sim 10^{-27}$ cm$^3$/s \cite{2015PhRvD..91l2002A}. As shown in figure \ref{fig:sv} we achieve this level of sensitivity for all DM masses up to $m_\chi \approx 500$ GeV. Thus we can conclude that the addition of the future GAMMA-400 high-quality data to the existing Fermi-LAT data significantly increases the sensitivity to the diphoton annihilation cross section (up to factor of 2, at least for the steep density profiles) and may reveal DM at any mass in the considered range $\approx$ (1--500) GeV.  

The sensitivity being estimated appeared to depend strongly on the fractional flux systematic uncertainty $\delta f_s$ mentioned above. As a natural basic scenario we assumed $\delta f_{sG} = \delta f_{sF} = 0.015$. To study the sensitivity dependence on this parameter we also calculated the sensitivity gains for the optimistic case of reduction of our systematics down to $\delta f_{sG} = 0.01$. The latter would yield the following gains with respect to Fermi-LAT by 12 years: 1.3--2.1 times for GAMMA-400 by 2 years alone and 2.1--2.7 times for GAMMA-400 by 4 years + Fermi-LAT by 12 years. Thus such potential optimization of systematics would provide an additional significant growth of sensitivity.

And finally we calculated GAMMA-400 sensitivity to the decaying DM for the similar exposures - the result is shown in figure \ref{fig:tau}. Here expectedly the sensitivity gain is not so large due to our limited sky coverage: GAMMA-400 by 2 years alone overtakes Fermi-LAT up to $m_\chi \approx 40$ GeV. However the combined analysis with 4 years of GAMMA-400 data will still provide the sensitivity gain by 1.1-2.0 times for the whole mass range! In this case the gain monotonically increases with the mass decrease.

\section{ALP discovery potential by an observation of the supernova explosion in the Local Group}
\label{sec:sn}
As was described in \cite{2017PhRvL.118a1103M}, gamma-ray telescopes has a high potential to discover ALPs by an observation of a nearby supernova (SN) explosion. In this section we aimed to estimate the GAMMA-400 sensitivity to photon-ALP coupling constant $g_{a\gamma}$ depending on the ALP mass $m_a$ by such an observation. Here we largely followed the methodology developed in \cite{2017PhRvL.118a1103M}. ALPs can be copiously produced in SN interiors during the explosion. Then they freely leave an SN, propagate and partially convert to photons in ambient magnetic fields. This may cause a gamma-ray burst from the SN during its explosion simultaneously with the neutrino burst. Such a gamma-ray burst typically peaks at 50--100 MeV \cite{2015JCAP...02..006P}, which fits well into the GAMMA-400 operating energy range. The first key step in this section was to model the expected spectrum of the burst. The time-integrated spectral flux has the form:
\begin{equation}
	\frac{dN_{\gamma}}{dE} = \frac{dN_{a}}{dE} \frac{P_{a\rightarrow\gamma}(E,m_a,g_{a\gamma},\vec{r}_{SN})}{4\pi d^2},
	\label{eq:sp-g}
\end{equation}
where $\frac{dN_{a}}{dE}$ is the energy distribution of the produced ALPs, $P_{a\rightarrow\gamma}(E,m_a,g_{a\gamma},\vec{r}_{SN})$ is their conversion probability on the way to observer and $d$ is the distance to SN. Let us write out the former two based on \cite{2017PhRvL.118a1103M} and \cite{2012PhRvD..86g5024H}:
\begin{equation}
	\frac{dN_{a}}{dE} = C_a \left(\frac{g_{a\gamma}}{\mbox{GeV}^{-1}}\right)^2 \left(\frac{E}{E_a}\right)^{\beta_a} \exp\left(-\frac{(\beta_a+1)E}{E_a}\right).
	\label{eq:sp-a}
\end{equation}
The constants $C_a,E_a,\beta_a$ moderately depend on the progenitor mass (see table 1 in \cite{2017PhRvL.118a1103M}). However, as was shown in \cite{2017PhRvL.118a1103M}, the final limits on $g_{a\gamma}$ depend on the progenitor mass very weakly. For this reason and the fact that the progenitor mass distribution is very steep we decided to fix $C_a,E_a,\beta_a$ to the following constant values, which reflect the most probable case of (10--20)$M_\odot$ progenitor: $C_a = 8 \cdot 10^{75}~\mbox{GeV}^{-1}, E_a = 99~\mbox{MeV}, \beta_a = 2.2$. The conversion probability in \eqref{eq:sp-g} has the form (in the natural units $c=\hbar=k_B=1$, see Eqs. (4)--(8) in \cite{2012PhRvD..86g5024H}):
\begin{equation}
\begin{aligned}
	P_{a\leftrightarrow\gamma}(E,m_a,g_{a\gamma},\vec{r}_{SN}) &= \frac{1}{1+(E_c/E)^2} \sin^2\left(\frac{g_{a\gamma} B_\bot d}{2}  \sqrt{1+\left(\frac{E_c}{E}\right)^2} \right),\\
	E_c &\simeq 2.5 \frac{|m_a^2-\omega_{pl}^2|}{\mbox{neV}^2} \left(\frac{\mu\mbox{G}}{B_\bot}\right) \left(\frac{10^{-11}~\mbox{GeV}^{-1}}{g_{a\gamma}}\right) ~\mbox{GeV}
	\label{eq:P}
\end{aligned}
\end{equation}
for the case of a uniform transverse magnetic field between the SN and observer. In reality the field is indeed not uniform. However, as will be shown below, the uniform field approximation can yield reasonably precise sensitivity estimates.

\begin{figure}[h]
\centering  
\includegraphics[width=0.9\textwidth]{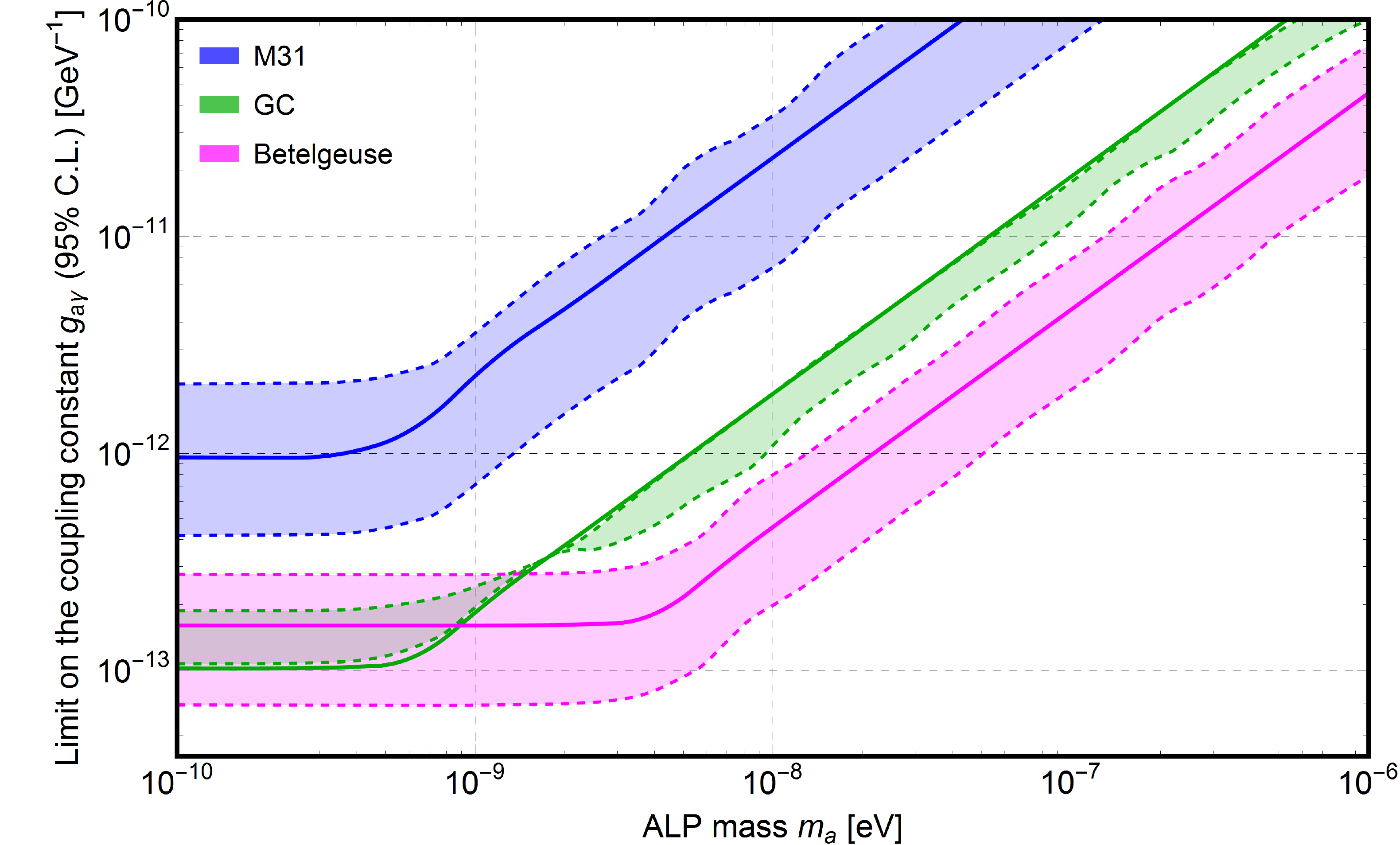}
\caption{\label{fig:sn-f} Fermi-LAT sensitivity to the ALP-photon coupling constant. The shaded areas reflect uncertainties in the limits due to uncertainties in the magnetic field distribution along the line of sight according to \cite{2017PhRvL.118a1103M}. The solid lines reflect the sensitivity in our simplified constant-field model with the chosen field values cited in table \ref{tab:sn}. See more details in section \ref{sec:sn}. }
\end{figure}

We calculated our sensitivity for three specific representative targets: SN at GC, in M 31 and Betelgeuse. These targets were also explored in \cite{2017PhRvL.118a1103M} for Fermi-LAT. As was outlined there, the magnetic field distribution uncertainties influence the limits on $g_{a\gamma}$ significantly: by factor of $\approx 2$ for the case of GC and by 4--5 for M 31 / Betelgeuse. These uncertainty ranges are shown in figure \ref{fig:sn-f} by the shaded areas bounded by the dashed lines. We found that such large ranges can easily enclose the sensitivity curves calculated with the simplified models of the uniform transverse magnetic field. This reduces a necessity to conduct a complicated procedure of numerical solving of the photon-ALP beam transport equation with the coordinate-dependent field. With the constant field model one can calculate the gamma-ray spectrum \eqref{eq:sp-g} analytically with the exactly known dependence on all the parameters. We decided to employ such an advantage and chose the simplified uniform field model for all our calculations. For this model we had to determine the effective transverse field value between the target and observer and the effective ALP path length. For the cases of GC and Betelgeuse the path lengths were naturally set to distances to these targets (see all the parameter values in table \ref{tab:sn}). For M 31 the path length is less obvious, since the major intergalactic portion of the propagation path practically does not contribute into the conversion probability. Hence we estimated the path length as the sum of portions inside Milky Way and M 31. For this we approximated the galaxies by cylinders with half-heights equal to the scale heights of their magnetic field distributions. These scale heights were taken from \cite{2013PhRvD..88b3504E,2016JCAP...03..060E}. Finally this yielded the effective path length $l_{eff} \approx 14$~kpc. Having set path lengths we found the effective transverse field values $B_{\bot eff}$, which yield approximately the average sensitivity of Fermi-LAT with respect to the boundary cases of field models reflected by the dashed lines in figure \ref{fig:sn-f}. To do this we used \eqref{eq:sp-g}--\eqref{eq:P}, Fermi-LAT characteristics and the data from table 2 in \cite{2017PhRvL.118a1103M}. The fitted field values are shown in table \ref{tab:sn} and respective $g_{a\gamma}$ Fermi-LAT limits are shown in figure \ref{fig:sn-f} by the solid lines. We see that these lines indeed fit rather well into the uncertainty ranges. Thus we consider our uniform field model as a model yielding middling conservative constraints with uncertainties about factor of 2. In other words, our model would reflect approximately an average sensitivity between the extreme cases. In general, the obtained field values are reasonable for the galactic medium (see e.g. \cite{2013PhRvD..88b3504E,2016JCAP...03..060E}). The field value for Betelgeuse is naturally much lower than that for the GC, since the former is located near the Galactic anticenter.
\begin{table}[h]
\caption{The parameters of three considered trial cases of SN explosion observations.}
\label{tab:sn}
\begin{center}
\begin{tabulary}{1\textwidth}{|C|C|C|C|}
\hline
 & Betelgeuse & GC & M 31 \\
\hline
Galactic longitude, deg & 200 & 0 & 121 \\
Galactic latitude, deg & -8.96 & 0 & -21.6 \\
Effective beam path $l_{eff}$, kpc & $\equiv d_{Bet} \approx 0.20$ & $\equiv d_{GC} \approx 8.4$ & 14 \\
Effective transverse magnetic field $B_{\bot eff},~\mu$G & 3.0 & 8.0 & 4.0 \\
Expected number of background counts $\mu_b$ & 0.2 & 1.0 & 0.1 \\
95\% limit on the number of signal counts $\mu_{sig,UL}$ & 3.0 & 4.2 & 3.0 \\
Probability to catch SN for GAMMA-400, \% & ? & 3--6 & 1--2 \\ 
\hline
\end{tabulary}
\end{center}
\end{table} 

At this point we determined all the ingredients for calculation of the gamma-ray spectrum. The next big step was to deduce the GAMMA-400 sensitivity to such a spectrum depending on $g_{a\gamma}$ and $m_a$. We integrated the spectrum over the energy range (50--400) MeV, which was found to be optimal for all the targets and is slightly narrower than in \cite{2017PhRvL.118a1103M}. Similarly to \cite{2017PhRvL.118a1103M} we set the signal integration time to be $\tau$ = 20 s. To obtain the achievable observational upper limit on the mean number of signal photons ($\mu_{sig}$) during this time interval in the presence of background we constructed the following simple Poissonian likelihood function:
\begin{equation}
	L_{SN}(\mu_{sig}(g_{a\gamma},m_a)|n,\mu_b) \propto \frac{(\mu_{sig}+\mu_b)^n}{n!}\exp(-\mu_{sig}-\mu_b),
\end{equation}
where $\mu_b$ is the mean number of background photons and $n$ is the measured counts. Similarly to the previous section \ref{sec:l} we assumed here $n = \mu_b$ rounding $\mu_b$ to the nearest integer (i.e. no apparent signal). Systematic flux uncertainties ($\sim$ 1\%) are irrelevant here. The mean photon numbers were calculated by the following way:
\begin{eqnarray}
	\mu_{sig} \equiv \mu_{sig}(g_{a\gamma},m_a) &&= 0.68\int\limits_{50MeV}^{400MeV} \frac{dN_{\gamma}}{dE}(E,g_{a\gamma},m_a) A_{eff}(E) dE, \label{eq:mssn} \\
	\mu_b &&\approx \int\limits_{50MeV}^{400MeV} f_b(E,b,l,\Omega_{PSF}(E=70\mbox{MeV})) A_{eff}(E)\tau dE. \label{eq:mbsn}  
\end{eqnarray}
The coefficient 0.68 in \eqref{eq:mssn} reflects the choice of ROI size as the size of PSF at 68\% containment. $f_b$ in \eqref{eq:mbsn} is the background spectral flux inside the ROI, which depends on a sky position and ROI/PSF size. For the latter we chose PSF at 70 MeV, since the spectrum \eqref{eq:sp-g} typically peaks around this energy. In general, $f_b$ includes all the components of gamma-ray sky. For our estimates we included only the diffuse backgrounds (Galactic and isotropic, from \cite{Fermi-map} using \cite{2000A&AS..143...33B}), since the contribution of point sources is expected to be subdominant (see e.g. figure 1 in \cite{2017ApJ...840...43A}). The obtained backgrounds are provided in table \ref{tab:sn}. We see that only the GC has rather significant background. Betelgeuse and M 31 can be considered as essentially background-free cases. And we neglected by the photon energy dispersion effect calculating \eqref{eq:mssn}-\eqref{eq:mbsn}, since it was checked to be irrelevant for the case of integration over so wide energy range. 

\begin{figure}[h]
\centering
\includegraphics[width=1\textwidth]{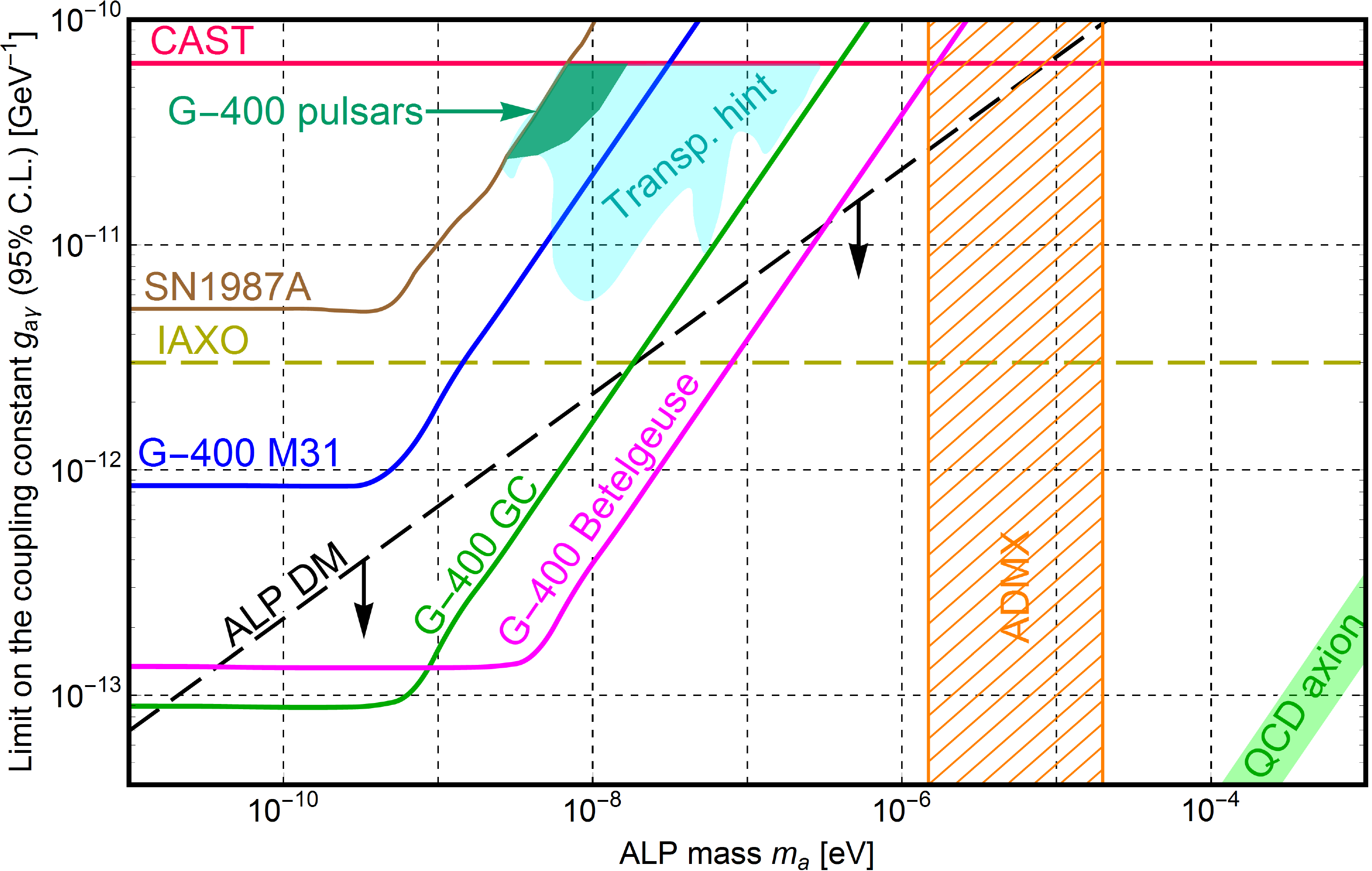}
\caption{\label{fig:sn} GAMMA-400 (G-400) sensitivity to the ALP-photon coupling constant by the observation of supernova in the GC, M 31; or Betelgeuse explosion. Other limits are also shown for comparison: CAST \cite{2018qchs.confE..34D}; future IAXO in its best configuration \cite{2018qchs.confE..34D}; absence of the gamma-ray burst from SN1987A \cite{2017PhRvL.118a1103M}; ADMX \cite{2015PNAS..11212278R}, which reaches the QCD axion band (outside of the plot range). See more details in section \ref{sec:sn}. Also the sensitivity region by observations of pulsars is shown, which overlaps with the anomalous TeV transparency hint region - see details in section \ref{sec:alp}. }
\end{figure}

To obtain the 95\% C.L. upper limit on the mean number of signal photons $\mu_{sig,UL}$ we implied the following condition analogically to \eqref{eq:0.95}:
\begin{equation}
	\int\limits_0^{\mu_{sig,UL}(g_{a\gamma}^{lim},m_a)} L_{SN}(\mu_{sig}|n=[\mu_b],\mu_b) d\mu_{sig} = 0.95.
\end{equation}
The obtained upper limits are presented in table \ref{tab:sn}. Then we substituted the obtained $\mu_{sig,UL}$'s in \eqref{eq:mssn} and found numerically the limiting coupling constant as a function of ALP mass $g_{a\gamma}^{lim}(m_a)$. The resulting curves are presented in figure \ref{fig:sn} together with other constraints for comparison. Overall we see that the obtained GAMMA-400 sensitivity is very similar to that of Fermi-LAT, which is shown by the solid lines in figure \ref{fig:sn-f} (for the same ALP propagation model). This is quite expected due to a comparable performance of both telescopes at the relevant energies. From figure \ref{fig:sn} we deduce that the sensitivity curves for GC and Betelgeuse do not differ drastically. For other locations of the SN in the Galactic disk we can expect intuitively the sensitivity curve lying somewhere between the curves for GC and Betelgeuse. Thus for $m_a \lesssim (10-100)$ neV an observation of the Galactic SN represents the most sensitive ALP search experiment among all existing. This would allow to explore completely the region of parameter space needed to explain the tentative anomalous transparency of the Universe at very high energies by ALPs \cite{2018qchs.confE..34D}. Also this observation would probe the certain portion of parameter space below the black dashed line in figure \ref{fig:sn}, where ALPs can accommodate all DM. It realizes for the mass range $m_a \sim (0.1-100)$ neV. However, it is still very hard to reach the well-motivated QCD axion parameter band, which is shown by green and represents the band enclosed between DFSZ and KSVZ axion models \cite{2018qchs.confE..34D}. So far this band has been reached only by ADMX experiment at $m_a \sim (1-10)~\mu$eV.

A non-trivial question here is indeed the probability to catch such an SN for GAMMA-400. The probability is the biggest for the case of GC, since as was mentioned in section \ref{sec:l}, according to the preliminary observational program we are going to observe the GC during 2--4 years. Let us estimate the probability of the catch based on the average Galactic rate of core-collapse SNe. The latter was thoroughly estimated in \cite{2013ApJ...778..164A} to be $\approx$ 3 SNe per century. We can naturally expect that mainly this rate is ``localized'' inside the Galactic disk. GAMMA-400 field of view centered on the GC captures $\approx$ 40\% of the whole Galactic disk. Counting that the star formation in the Galaxy is slightly concentrated towards the center, it would be reasonable to assume that at least $\approx$ 3/2 SNe/century can be expected in average for GAMMA-400 looking at the GC. Considering the SNe explosions to be a Poissonian process, one can easily get that the probability of one or more SNe during 2--4 years of quasi-continuous observations is 3--6\%. According to preliminary plans GAMMA-400 will also observe a significant part of the Galactic plane beyond the GC region during $\approx$ 1.5 years. In such a case this would add another $\approx$ 1\% of probability to the total chance. M 31 is expected to be inside the field of view during 0.4--0.7 years, which adds 1--2\%. We do not really consider the Magellanic clouds and M 33 due to very low chances from them. Thus the total chance is estimated to be 5--9\%. Besides that we may keep in mind two famous supergiants, which are very close to the explosion - $\eta$ Car Foramen and $\alpha$ Ori Betelgeuse. The latter is especially interesting due to its proximity, which would allow to predict the explosion several days in advance thanks to the pre-supernova steady neutrino signal \cite{2004APh....21..303O}. This would provide a great opportunity to point GAMMA-400 at Betelgeuse in advance and observe the explosion with a guarantee in case it will happen during the telescope lifetime. In general, as was also noted in \cite{2017PhRvL.118a1103M}, the neutrino signal accompanying the explosion is very useful timing trigger for searches of the gamma-ray burst being considered. Neutrino telescopes will easily detect the neutrino burst from any Galactic SN. However, M 31 in this aspect is less favorable being too far for the neutrino burst to be detectable.

And finally we would like to discuss briefly potential uncertainties of the sensitivity estimated here. For this purpose we investigated the dependence of the limiting coupling constant value on the key parameters and obtained approximately the following for the relevant values of other parameters: $g_{a\gamma}^{lim} \propto B_\bot^{-1/2} \mu_{sig,UL}^{1/4}$. Thus the dependence on the number of signal photons is very weak and on the magnetic field is mild. Above we mentioned that we neglected by the contribution of point sources in $\mu_b$. This approximation may lower $\mu_b$ and, hence, $\mu_{sig,UL}$. However $g_{a\gamma}^{lim}$ practically would not change even if we would double or triple $\mu_b$ - due to so weak dependence on $\mu_{sig,UL}$. Thus our sensitivity estimate is quite realistic with respect to the background model. As we already mentioned, the main source of uncertainty is the magnetic field model. We would estimate the respective systematic uncertainty of $g_{a\gamma}^{lim}$ to be at the level of factor of $\approx$ 2. Meanwhile, we also studied two possibilities to detect the gamma-ray burst - by the whole converter and only by its thin part. The latter has much smaller PSF (see details in section \ref{sec:i}), which would reduce the background. However, we found that the sensitivity of both variants is approximately the same due to the reduced effective area in the ``thin'' mode. Another advantage could be obtained from using the lateral aperture mentioned in section \ref{sec:i}. Together with the main aperture the total field of view may reach $\approx$ 1/3 of the whole sky. This significantly increases the probability of the catch estimated above for the main aperture by $\approx$ 3\% bringing the total chances to 8--12\%. The lateral aperture will presumably have a large background level. However the sensitivity to $g_{a\gamma}$ may still be at the relevant level due to very weak dependence of the former on $\mu_{sig,UL}$. Overall, although we are not ready to really quantify exact predictions for the lateral aperture, we consider it to be a useful mode for the detection of SNe and other transients.

\section{ALP constraining potential by observations of bright gamma-ray pulsars}
\label{sec:alp}
Bright gamma-ray pulsars located near the Galactic plane represent a good target for ALP searches by GAMMA-400. For certain values of ALP mass and photon coupling constant a significant part of photons can convert to ALPs during propagation in the Galactic magnetic field forming the oscillations in the pulsar spectra according to \eqref{eq:P}. Galactic pulsars are generally better than AGNs for ALP searches due to much better knowledge of the magnetic field distribution between the source and observer. In \cite{2018JCAP...04..048M} an attempt was made to find such oscillations based on Fermi-LAT data and resulted in a tentatively positive detection of ALP with $m_a \approx$ 3.6 neV, $g_{a\gamma} \approx 2.3 \cdot 10^{-10}$ GeV$^{-1}$. In this section we conducted the estimation of GAMMA-400 sensitivity to such oscillations and respective ALP parameters. In general, one can intuitively expect a comparable sensitivity of GAMMA-400 and Fermi-LAT to ALPs. However better angular and energy resolutions of GAMMA-400 (figure \ref{fig:char}) may enable better sensitivity by decreased background subtraction systematics and spectral smearing. In our estimates we decided to study the sensitivity on the ``$m_a - g_{a\gamma}$'' parameter plane outside of the regions excluded by CAST and non-observation of gammas from SN1987A (see figure \ref{fig:sn}). We used the same 6 pulsars employed by \cite{2018JCAP...04..048M} (PSRs J1420-6048, J1648-4611, J1702-4128, J1718-3825, J2021+3651, J2240+5832) as well-suitable for our purposes: they are bright to provide good photon statistics, distant to gain a significant ALP-photon conversion probability and located very close to the Galactic plane, where the magnetic field is the strongest. Only one modification to the pulsar sample we made is the assumed distance to PSR J2021+3651. \cite{2018JCAP...04..048M} assumed the distance of 10 kpc. However the dedicated study of this pulsar \cite{2015ApJ...802...17K} emphasizes a generally big uncertainty in the distance, which can be in the range 1--12 kpc. Their additional study allowed to constrain the distance to the range $1.8_{-1.4}^{+1.7}$ kpc. In such a situation we decided that 10 kpc would be too optimistic (for our purposes) assumption. Also for such a distance the pulsar appear suspiciously bright in gamma rays. Thus we decided to stay conservative and placed it at the upper distance obtained in \cite{2015ApJ...802...17K}, i.e. 3.5 kpc. Distances to other pulsars were taken from \cite{2018JCAP...04..048M}.

\begin{figure}[h]
\centering
\includegraphics[width=0.9\textwidth]{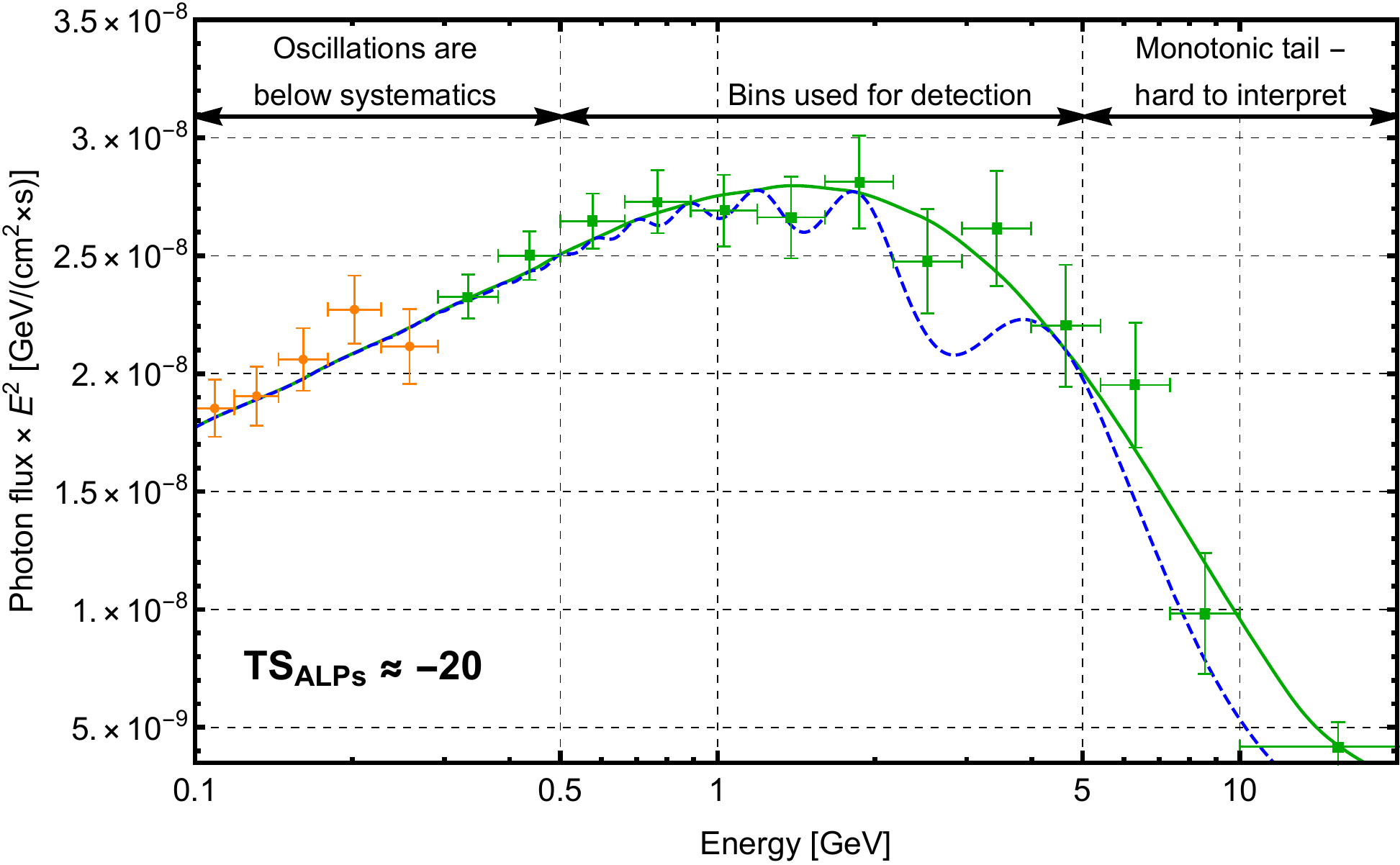}
\caption{\label{fig:alp} The spectra of PSR J1420-6048 with (blue dashed) and without (green) oscillations caused by ALPs ($m_a$ = 7.0 neV, $g_{a\gamma} = 4.6 \cdot 10^{-11}$ GeV$^{-1}$) together with the simulated GAMMA-400 data points. The exposure time is 0.5 years. The orange data points at low energies reflect the usage of the thin converter. More details are in section \ref{sec:alp}. }
\end{figure}

Our algorithm of sensitivity estimation is based on the likelihood function, which joins together all the pulsars and all the relevant energy bins of each pulsar. Briefly the steps of our procedure are the following: we
\begin{enumerate}
	\item Took the intrinsic spectra of pulsars, which were obtained in \cite{2018JCAP...04..048M} (red dashed lines in their figures with spectra) and are supposed to be cleaned out of instrumental systematics and modulations due to ALPs.
	\item Chose an optimal binning for our telescope (can be seen in figure \ref{fig:alp}) and simulated the data points assuming no ALPs and taking into account both the statistical and systematical fluctuations (the latter was described in section \ref{sec:l}).
	\item Chose a particular ($m_a, g_{a\gamma}$) values being tested and generated the spectra of pulsars with ALPs by multiplication of the intrinsic spectra by the probability of photon-ALP conversion \eqref{eq:P} as function of energy.
	\item Constructed the following joint likelihood function:
	\begin{equation}
		L(m_a,g_{a\gamma}) \propto \prod_{j=1}^{6PSRs} \prod_{i=1}^{\#bins_j} L_{ij}(\mu_{ij}(m_a,g_{a\gamma}),n_{s,ij}|n_{ij}),
		\label{eq:La}
	\end{equation}
	where $L_{ij}$ is the individual likelihoods for each bin. It has the form similar to \eqref{eq:L} with $n_{s,ij}$ being systematic offsets, $n_{ij}$ - measured (i.e. simulated at step 2) numbers of photons and $\mu_{ij}$ - the expected (average) numbers of photons:
\begin{equation}
\begin{aligned}
		\mu_{ij}(m_a,g_{a\gamma}) = \int\limits_{E_{il}}^{E_{iu}} dE' \int dE f_j(E) &\times \left\{
		\begin{array}{ll}
		1~\mbox{for null likelihood}\\
		1-P_{\gamma\rightarrow a}(E,m_a,g_{a\gamma},B_{\bot eff},d_j)
		\end{array} 
		\right\} \times \\ &\times
		\frac{1}{\sqrt{2\pi}\sigma_E(E)}\exp\left(-\frac{(E-E')^2}{2\sigma_E^2(E)}\right)\varepsilon_j(E),
\end{aligned}
\label{eq:mua}
\end{equation}
where $E_{il},E_{iu}$ are the energy bin margins, $f_j(E)$ are the intrinsic photon spectra of pulsars and $d_j$ are the pulsar distances.
	\item Constructed the criterion of rejection of the hypothesis of ALPs presence in the spectra through the standard test statistic approach:
	\begin{equation}
		TS_{ALPs}(m_a,g_{a\gamma}) = 2\ln \frac{L(m_a,g_{a\gamma},\vec{n}_s^{max})}{L(m_a=0,g_{a\gamma}=0,\vec{n}_{s0}^{max})} \equiv 2(\ln L - \ln L_0),
	\end{equation}
where $\vec{n}_s^{max},\vec{n}_{s0}^{max}$	vectors of nuisance systematic offsets are fitted in each bin to maximize $L$ and $L_0$ respectively. This approach is actually different from the one used in section \ref{sec:l}, where we marginalized the likelihood over the nuisance parameters. Here the latter approach is practically inconvenient, so we decided to employ the likelihood ratio test approach, which is also widely used and often called the frequentist approach (see e.g. \cite{2020JCAP...02..012H} for a general methodological review). Then we set the following typical model rejection criterion (see also \cite{2014JCAP...10..023A}): $TS_{ALPs}(m_a,g_{a\gamma}^{lim}) = -2.71$. Physically by construction it means that for each ALP mass $m_a$ the coupling constant values $g_{a\gamma} \geqslant g_{a\gamma}^{lim}$ are excluded with the confidence level $\geqslant$ 95\%. In other words, the spectral oscillations produced by $g_{a\gamma} \geqslant g_{a\gamma}^{lim}$ would be large enough to be inevitably distinguishable from the smooth spectra without ALPs.
	\item Repeated steps 3--5 for all relevant discrete pairs of values ($m_a,g_{a\gamma}$) and obtained the interpolated mean sensitivity exclusion contour $g_{a\gamma}^{lim}(m_a)$ at 95\% confidence level. The obtained sensitivity region is drawn by the aquamarine color in figure \ref{fig:sn}.
\end{enumerate}

Let us discuss several essential details about the procedure outlined above. The conversion probability $P_{\gamma\rightarrow a}$ determined by \eqref{eq:P} and participating in \eqref{eq:mua} indeed requires the specification of magnetic field between the pulsars and observer. Here similarly to section \ref{sec:sn} we set the uniform field, which was shown there to provide approximately the medium sensitivity among various possible field models. We set $B_{\bot eff} = 5~\mu G$ for all the pulsars as an average value between those obtained in section \ref{sec:sn} as representative values for the directions towards the Galactic center and anticenter (the cases of GC and Betelgeuse in table \ref{tab:sn} respectively). Such choice is motivated by the fact that all the participating pulsars lie far from both the center and anticenter - their average angular distance from the GC is $\approx50\degree$. As for the assumed effective exposure times, they vary between 0.25 and 1.8 years depending on the pulsar location. The pulsars located closer to the GC get longer exposure times - they will be captured in the field of view during the preliminary planned deep 2+ year observations of the GC. In order to calculate the mean sensitivity realistically we simulated 1000 random data samples $n_{ij}$ and took those $g_{a\gamma}^{lim}(m_a)$, which yields $\langle TS_{ALPs}(m_a,g_{a\gamma}^{lim}) \rangle = -2.71$ with the latter being averaged over the ensemble of samples. 

Figure \ref{fig:alp} shows the example of particular pulsar spectra with respective data points for the case of ($m_a,g_{a\gamma}$) values, which are approximately central for the sensitivity region in figure \ref{fig:sn}. The non-trivial question for the procedure above was - which bins should we include in the likelihood \eqref{eq:La}? Figure \ref{fig:alp} illustrates the choice of bins: we did not include the bins at low energies where $P_{\gamma\rightarrow a} \lesssim 0.01$, since the oscillations there would be indistinguishable from systematics; and the bins at high energies where the coupling is strong, and $P_{\gamma\rightarrow a}$ does not oscillate anymore increasing monotonically. Such a high energy ``tail'' was not included in the analysis simply because it is unclear how to distinguish between the intrinsic spectral slope, which can be not known exactly a priori, and the possible absorption by ALPs at these energies. Thus the obtained sensitivity is rather conservative in this respect. The first 5 bins in the energy range of 100--300 MeV were generated by the thin converter for all the pulsars, since the whole converter has large PSF at these energies and, hence, will inevitably intrude large background systematics in the data. In general, by looking at figure \ref{fig:alp} one can not see that the data points (i.e. 8 bins used for detection) prefer the smooth spectrum really obviously. However the obtained sensitivity is achieved by stacking together the multiple bins and pulsars. From figure \ref{fig:sn} the sensitivity reaches $g_{a\gamma}^{lim,min} \approx 2.4 \times 10^{-11}~\mbox{GeV}^{-1}$ at $m_a \approx 3$ neV and then worsens as $m_a$ both decreases and increases up to the CAST limit at $m_a \sim$ 1, 10 neV. This reproduces a typical sensitivity mass range for the space-based gamma-ray telescopes, see e.g. \cite{2016PhRvL.116p1101A}. With increase of $m_a$ the spectral oscillations generally shift to higher energies.

Meanwhile our estimated sensitivity is by an order of magnitude better than the mentioned coupling constant value obtained in \cite{2018JCAP...04..048M} as a tentative signal. We can not compare it with our sensitivity directly due to different magnetic field model, which indeed influences $P_{\gamma\rightarrow a}$ and, hence, the sensitivity. However we suppose that $g_{a\gamma} \approx 10g_{a\gamma}^{lim}$ would produce big oscillations detectable by GAMMA-400 for any reasonable case of magnetic fields, especially taking into account an opportunity of the joint data analysis of GAMMA-400 and Fermi-LAT. Thus we anticipate to confirm or deny robustly the tentative detection \cite{2018JCAP...04..048M} by GAMMA-400, even if such ALP evaded by somehow CAST and SN1987A constraints. If the detection will not be confirmed, then the joint data analysis from both telescopes may expand the GAMMA-400 sensitivity region further and test a major part of the blue region in figure \ref{fig:sn}, which reflects the parameter values needed to fit the anomalous transparency of the Universe at very high energies by ALPs. Estimation of the joint sensitivity of both telescopes requires a bit of extrapolation and is left for a future work. Also the GAMMA-400 sensitivity region can be further expanded by an increase of the pulsar exposures, which were assumed here to be minimally anticipated and can be significantly larger in the optimistic case of telescope operation during 10 years. Indeed our estimates here are based on the simplistic uniform magnetic field model. But in this respect we would like to note that in order to study properly the systematical uncertainties related to the field model, it is not enough to model realistically the Galactic magnetic field only like it was done in \cite{2017PhRvL.118a1103M,2018JCAP...04..048M}. One also has to evaluate the potential influence of the intrinsic/PWN and heliomagnetic fields, which inevitably exist along the line of sight. Finally, the sensitivity can indeed be further increased by adding other objects into the sample including supernova remnants \cite{2019PhRvD.100l3004X}.

\section{Miscellaneous targets}
\label{sec:misc}
\begin{figure}[h]
\centering
\includegraphics[width=0.8\textwidth]{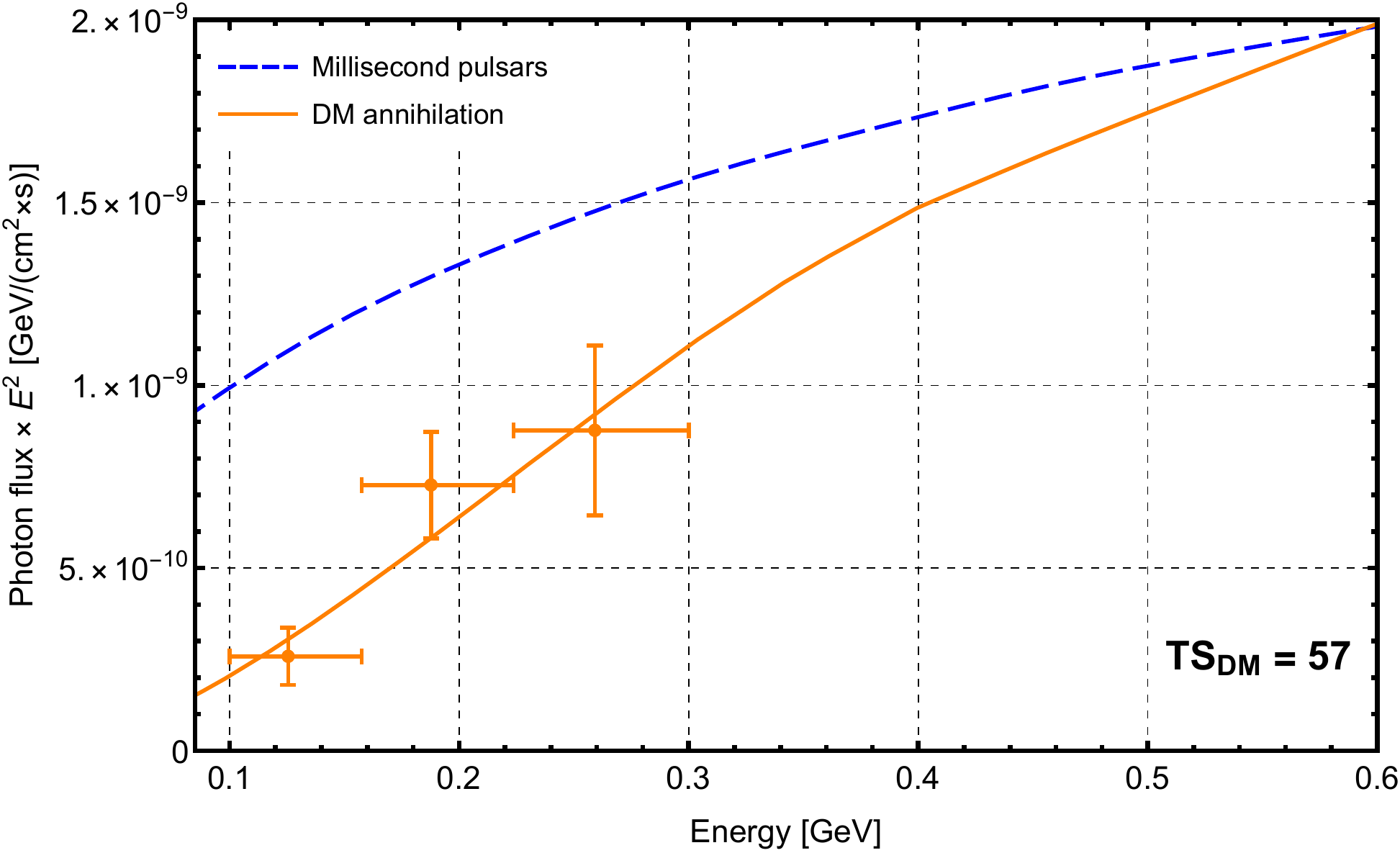}
\caption{\label{fig:om} The fitted spectra of the globular cluster $\omega$ Cen assuming millisecond pulsars and annihilating DM as the sources of emission (from \cite{2019arXiv190708564B}) together with the simulated GAMMA-400 data points at low energies. Two spectra can be reliably distinguished. The assumed exposure time is 0.7 years.}
\end{figure}

\textbf{\textit{Globular clusters.}} Recently the detection of gamma rays from the globular clusters 47 Tuc and $\omega$ Cen was reported, and DM annihilation as one of the possible emission sources was proposed \cite{2018PhRvD..98d1301B,2019arXiv190708564B}. However soon after publication \cite{2019arXiv190708564B} 5 millisecond radio pulsars were discovered in $\omega$ Cen \cite{2020ApJ...888L..18D}. The latter makes pulsars to be much more probable gamma-ray source than DM annihilation. However still considering both options we estimated that GAMMA-400 will be potentially able to distinguish robustly these two sources employing a good angular resolution at low energies. Figure 3 in \cite{2019arXiv190708564B} shows the fits of the $\omega$ Cen spectrum (measured by Fermi-LAT) by pulsars and annihilating WIMPs. From this figure one can see that both possible sources have very similar spectra above $\approx$ 300 MeV, which makes them hard to disentangle. Below this energy the spectra differ significantly. But over there Fermi-LAT does not really see the object presumably due to large systematics, which are caused by large PSF (even for Fermi-Front) and, hence, high background contamination from the nearby Galactic plane etc. Thus better measurements at low energies may provide a crucial information to finally determine the emission mechanism. This information could be gathered employing the thin converter of GAMMA-400, which has the PSF $\approx$ 2 times smaller than that of Fermi-Front and, hence, smaller background systematics. Assuming the spectrum of DM, we simulated the idealized (i.e. no background systematics) GAMMA-400 data points for $\omega$ Cen at low energies - see figure \ref{fig:om}. We see quite perfect distinction between two possible sources with $TS_{DM} = 57$. Although it is generally unlikely to find DM in this object, this is a good example, which demonstrates the advantage of GAMMA-400 at low energies. 

\textbf{\textit{Subhalos and dwarfs}} continue to be interesting targets for DM searches in gamma rays. We provided some estimates of subhalo detectability by GAMMA-400 in \cite{2018PAN....81..373E}. As for the dwarf Milky Way satellites, although GAMMA-400 primarily is not designated for such faint high-latitude objects, potentially it may provide an improvement in the sensitivity to WIMPs annihilating there with the masses around $m_{\chi} \sim 100$ GeV. Such WIMP masses (assuming the thermal cross section) are poorly reached by both Fermi-LAT and ground-based Cherenkov telescopes including CTA, see e.g. figure 3 in \cite{2019PhRvD..99l3017H}. Thus additional observations by GAMMA-400 with an excellent quality of photons above 10 GeV may play a significant role in ``bridging'' this mass gap.  

\textbf{\textit{Galactic center}} historically has the well-known gamma-ray excess (e.g. \cite{2014PhRvD..90b3526A,2017ApJ...840...43A}). The debate on its origin is still actively going on with strong arguments favoring both opportunities: the annihilating DM origin \cite{2019PhRvL.123x1101L} and the origin from millisecond pulsars \cite{2020PhRvD.102d3012A}. GAMMA-400 again can help to resolve finally this puzzle adding a new high quality data by the planned deep observations of the GC. Especially big progress may come from low energies: the line of reasoning similar to the one developed above for $\omega$ Cen applies to the GC as well. Also besides traditional WIMPs another recently proposed peculiar DM candidate in the form of the Bose-Einstein condensate of the hexaquark d*(2380) particles \cite{2020JPhG...47cLT01B} has the best search prospects at sub-GeV energies in the GC as was studied in \cite{2020arXiv200309283B}!

\textbf{\textit{NGC 1275 and other galaxies.}} NGC 1275 is a gamma-ray bright galaxy at the center of the Perseus galaxy cluster, which is supposed to host high central magnetic fields. Hence this is a good target for ALP searches, which are actively going on \cite{2016PhRvL.116p1101A}. Although we plan to observe this object, we do not rely on it as much as on pulsars described in section \ref{sec:alp}. The reason is simply that the magnetic field distribution around NGC 1275 is less known, which produces big uncertainties in the ALP constraints derived from this object as was shown in the recent paper \cite{2020PhLB..80235252L}. Thus according to \cite{2020PhLB..80235252L} some alternative magnetic field models are able to elevate the limiting coupling constant values above the CAST limit (!), making the constraints from this target too model-dependent and, hence, ambiguous. Nevertheless, other similar sources - particularly, M 87 - can be studied too. Finally, the recent paper \cite{2019ApJ...880...95K} identified hints of annihilating WIMPs in the outer halo of M 31 in gamma rays. This makes M 31 a potentially interesting target for GAMMA-400 too.

\textbf{\textit{Other axion-related probes.}} Recently the paper \cite{2020arXiv200408785B} proposed another way to probe $\sim$meV-scale axions: they can be produced inside neutron stars, escape them and then subsequently decay into gamma photons, which could be detectable as an extended halo around the neutron star. This halo is expected to have the angular radius $\approx 2\degree$ at 10--200 MeV energies for the neutron star at distance $\sim$ 100 pc. Such halo could be resolved as an extended source by the thin converter of GAMMA-400 (see figure \ref{fig:char}). Other work \cite{2020JCAP...01..058G} points to the opportunities at low energies too: more exotic DM in the form of axino and/or gravitino can be detected through the spectral lines from their decay in the GC.

\section{Conclusions}
\label{sec:last}
We analyzed the capabilities of the planned space-based gamma-ray telescope GAMMA-400 for DM indirect detection. We mainly (but not only) focused on the two most popular DM candidates - WIMPs and ALPs. GAMMA-400 is expected to have improved angular and energy resolutions in comparison with Fermi-LAT, which would provide new opportunities in several directions of DM searches. The preliminary observational program of GAMMA-400 includes the deep pointed observations of the GC region during 2--4 years, deep survey of the selected parts of the Galactic plane during 1.5--2 years and other targets. Such program is expected to enable the following opportunities.

We estimated the mean sensitivity of GAMMA-400 to the narrow lines due to DM annihilation or decay in the GC and compared it with the estimated sensitivity of Fermi-LAT after 12 years of work assuming Einasto density profile for the range of DM particle masses (1--450) GeV ((1--900) GeV for decay). We obtained that with the mentioned 2--4 years exposure GAMMA-400 is more sensitive to the annihilating DM in the whole mass range and will be able to test the diphoton annihilation cross sections at least by 1.2--1.5 times smaller than that for Fermi-LAT by 12 years (figure \ref{fig:sv}). The joint analysis of Fermi-LAT and GAMMA-400 data will yield in the optimistic case the sensitivity gain up to a factor 1.8--2.3 depending on DM mass reaching $\langle \sigma v \rangle_{\gamma\gamma}(m_\chi = 100~\mbox{GeV}) \approx 10^{-28}$ cm$^3$/s. This in turn will allow to test comprehensively the hypothesized line at $E_\gamma = m_\chi = (64-67)$ GeV predicted by the specific DM model \cite{2019PhRvD.100c5023C}, as well as model \cite{2020JHEP...07..148Y} partially. This gives a good hope to reveal DM in case it is composed of those candidates. For the decaying DM the joint analysis may provide the gain of sensitivity to the DM lifetime up to a factor 1.1-2.0 reaching $\tau_{\gamma\nu}(m_\chi = 100~\mbox{GeV}) \approx 2 \cdot 10^{29}$ s (figure \ref{fig:tau}).

Another promising direction is the ALP search by a potential observation of SN explosion in the Local Group. ALPs are very popular DM candidate, can be produced and emitted in SN and then detected as gamma-ray burst due to conversion to photons during propagation. Such a burst typically peaks at 50--100 MeV and represents a very sensitive probe of ALP parameters. Thus we showed that the observation of Galactic SN by GAMMA-400 can constrain the photon-ALP coupling constant down to the level $g_{a\gamma} \sim 10^{-13}~\mbox{GeV}^{-1}$ for ALP masses $m_a \lesssim 1$ neV (figure \ref{fig:sn}). This represents the most sensitive indirect probe of ALPs, which allows to test some part of the ALP parameter space, where ALPs can explain all DM. For the case of SN in M 31 the sensitivity is lower by about an order of magnitude. We estimated our sensitivity in the frame of the simplified uniform magnetic field model. However, as we noted, the sensitivity dependence on the assumed magnetic field model is medium. The total chance to catch SN in the Local Group during the mission lifetime is $\approx$ 8--12\%.

ALPs can be also constrained by observations of bright and distant Galactic gamma-ray pulsars. ALPs with certain parameter values would produce noticeable modulations in the pulsar spectra due to photon-ALP conversion in the Galactic magnetic fields. We estimated the GAMMA-400 sensitivity to ALP parameters causing such modulations by the simulation of observations of 6 convenient pulsars. GAMMA-400 appeared to be more sensitive than CAST in the mass range $m_a \approx (1-10)$ neV reaching $g_{a\gamma}^{min} \approx 2 \cdot 10^{-11}~\mbox{GeV}^{-1}$ (figure \ref{fig:sn}). This will allow to robustly test the tentative ALP detection in the spectra of these pulsars by Fermi-LAT reported in \cite{2018JCAP...04..048M}.

Other potentially interesting targets for WIMP searches include: the Galactic center, globular clusters, subhalos and dwarf satellites. For ALPs these are AGNs, particularly NGC 1275 and M 87. These targets are briefly described in section \ref{sec:misc}.

Indeed further development of our activities is anticipated as the mission will be approaching to the launch. The directions of development are mainly elaboration of other possible targets for DM searches, especially those outlined in section \ref{sec:misc}; refinement of the physical models involved and keeping the mission objectives up-to-date according to the developments in the field. This paper aimed to provide the sensitivity estimates at the first approximation using the current level of knowledge in the gamma-ray astronomy. The obtained sensitivity of GAMMA-400 to various DM candidates allow us to expect a significant contribution of this planned telescope to the field of DM indirect searches.



\acknowledgments
For a major part of all our calculations we use Wolfram Mathematica package and would like to acknowledge its excellent performance for the scientific computing and visualization. Our research has made use of "Aladin sky atlas" developed at CDS, Strasbourg Observatory, France \cite{2000A&AS..143...33B} and the WebPlotDigitizer \cite{WPD}. This study was supported by the Russian State Space Corporation ROSCOSMOS (contract no. 024-5004/16/224).


\bibliography{C:/Users/Andrey/YandexDisk/DM/universal}

\end{document}